\documentclass[final,authoryear,5p,times,twocolumn]{elsarticle}

\usepackage{graphicx}

\usepackage{amssymb}
\usepackage{amsthm}
\usepackage{amsmath}

\usepackage{xcolor}

\usepackage{lineno}
\usepackage{verbatim}
\usepackage[normalem]{ulem}

\journal{Icarus}

\begin{document}

\renewcommand{\vec}[1]{\mathbf{#1}}
\newcommand{\rem}{}
\newcommand{\stephan}{}
\newcommand{\jan}{}
\newcommand{\stephannew}{}
\newcommand{\neu}[1]{#1}
\newcommand{\alt}[1]{}

\begin{frontmatter}


\author{Jan Verhoeven\fnref{label1}}
\fntext[label1]{Corresponding author.}
\ead{jan.verhoeven@uni-muenster.de}

\title{The compressional beta effect: a source of zonal winds in planets?}


\author{Stephan Stellmach}

\address{Westf\"alische Wilhelms-Universit\"at, Institut f\"ur Geophysik, Corrensstr. 24, 48149 M\"unster, Germany}

\begin{abstract}
Giant planets like Jupiter and Saturn feature strong zonal wind patterns on their surfaces. Although several different mechanisms that may drive these jets have been proposed over the last decades, the origin of the zonal winds is still unclear.
Here, we explore the possibility that the interplay of planetary rotation with the compression and expansion of the convecting fluid can drive multiple deep zonal jets by a compressional \neu{Rhines-type}\alt{beta} mechanism, as originally proposed by \citet{Ingersoll1982}. In a certain limit, this deep mechanism is shown to be mathematically analogous to the classical \neu{Rhines}\alt{beta} mechanism possibly operating at cloud level.  Jets are predicted to occur on a compressional Rhines length $l_R = (2 \Omega \langle H_\rho^{-1} \rangle v_{jet}^{-1} )^{-1/2}$, where $\Omega$ is the angular velocity, $\langle H_\rho^{-1} \rangle$ is the mean inverse density scale height and $v_{jet}$ is the typical jet velocity. Two-dimensional numerical simulations using the anelastic approximation reveal that this mechanism robustly generates jets of the predicted width, and that it typically dominates the dynamics in systems deeper than $O(l_R)$. Potential vorticity staircases are observed to form spontaneously and are typically accompanied by unstably stratified buoyancy staircases. The mechanism only operates at large rotation rates, exceeding those typically reached in three-dimensional simulations of deep convection in spherical shells.  Applied to Jupiter and Saturn, the compressional Rhines scaling reasonably fits the available observations. Interestingly, even weak vertical density variations such as those in the Earth core can give rise to a large number of jets, leading to fundamentally different flow structures than predicted by the Boussinesq models typically used in this context.  
\end{abstract}


\end{frontmatter}



\section{Introduction}
\label{introduction}

Strong zonal winds organize the colorful clouds on Jupiter's surface into banded structures. These cloud patterns have already been observed with telescopes more than 350 years ago \citep{Rogers1995}, and since then, scientists have studied them with ever more sophisticated observation tools. In addition to optical telescopes, the Hubble space telescope and the Pioneer, Voyager, Galileo and Cassini spacecraft missions have revealed fascinating pictures of the complex wind patterns shaping the surfaces of the gas and ice giants in our planetary system. Jupiter and Saturn exhibit strong prograde (i.e. eastward) equatorial jets, which are flanked by weaker, alternating westward and eastward winds on each hemisphere. Uranus and Neptune also exhibit pronounced zonal winds, but in contrast to Jupiter and Saturn, strong retrograde equatorial jets are observed.

It is still unknown how deep the winds extent into the interior. The Galileo probe that entered Jupiter's atmosphere down to $150$ km in 1995 gave evidence for an increase of the wind speeds at larger depth \citep{Atkinson1997}, but provided little information about the deep interior. The Juno mission that will reach Jupiter in mid-2016 is expected to better constrain the radial extent of the jets by carrying out high-resolution measurements of Jupiter's gravity field \citep{Kaspi2010}.

The theoretical understanding of the zonal winds is still incomplete. The different theories proposed so far are commonly classified into two distinct groups. The first class of models suspects the key to the zonal winds in a shallow layer at cloud level, with energy being pumped into the jets by processes like moist convection, lateral variations in solar heating or other processes occurring close to the surface. In contrast, the second class of models views the zonal winds as an expression of processes occurring deep in the planetary interior, typically driven by convective instabilities. In the following, we will refer to these two classes as shallow- and deep-forcing models. They represent end-member cases of potential forcing scenarios, and it is possible that a combination of both, deep- and shallow-forcing, is needed in order to explain all observational data.

It is instructive to reconsider the key processes driving zonal flows in the different approaches.  
The models following the shallow-forcing paradigm typically consider a fluid confined to a thin layer at the planetary surface, which leads to considerable simplifications of the governing equations. In the simplest cases, two-dimensional, incompressible flow is assumed, while more advanced approaches include shallow-water and multi-layer models. In all these cases, the latitudinal variation of the tangential component of the Coriolis force, the so-called {\em beta effect}, plays a key role, as it forces fluid parcels moved in latitudinal direction to change their vorticity. The effect becomes significant for large flow structures only, whereas the dynamics on small and intermediate scales is usually characterized by an inverse cascade of kinetic energy. This turbulent upward cascade ceases at the so-called Rhines length \citep{Rhines1975} when the beta effect becomes felt by the flow. From this length scale on, the flow dynamics is dominated by Rossby waves, which leads to a strong anisotropy of the large scales and ultimately to the formation of jets.  A large number of theoretical, experimental and numerical studies have confirmed the robustness of this now classical picture of zonal wind generation (see \citet{Vasavada2005} for a review). While earlier works typically report retrograde equatorial jets, more recent simulations have also succeeded in producing prograde equatorial jets by including additional physical processes like energy dissipation by radiative relaxation or latent heating resulting from the condensation of water vapor (e.g. \citet{Cho1996,Williams2003,Showman2004,Scott2008,Lian2010}). 

In contrast to these shallow models, in the deep-forcing scenario, convection in a fluid confined to a deep spherical shell is considered. In case of the gas giants, the inner boundary is often assumed to be set by the transition from molecular to metallic hydrogen, where Lorentz forces become important and are thought to lead to different flow dynamics deeper within the planet.
As in the shallow models, Rossby waves are believed to play a central role in driving deeply seated zonal jets as well. However, the processes of local vorticity generation that are responsible for the waves are governed by different physics. Mainly, two mechanisms have been identified as possible sources of Rossby waves in the deep interior -- the so-called {\em topographic beta effect} and a process that we call the {\em compressional beta effect} in this paper. Both are reviewed in more detail below.

 Apart from different wave mechanisms considered, theories on deep jet generation also differ in their mechanistic view on how these waves channel kinetic energy into zonal winds. Table \ref{mechanism_table} gives a schematic overview over several popular models, along with their underlying assumptions and key predictions.  Even though this table only presents a somewhat simplified view and fails to include all aspects of the various theories, we feel that it is nevertheless useful in providing a compact overview of the similarities and differences between the various approaches.  Note that these different views of the jet generation mechanics are complementary in many ways, and should not be seen as being necessarily contradicting. In the following, we give a brief review of the different models, which allows us to discuss the ``compressional beta effect'' studied in this paper in a broader context.

\begin{table*}[bt]
\label{mechanism_table}
\begin{center}
\includegraphics[width=\linewidth]{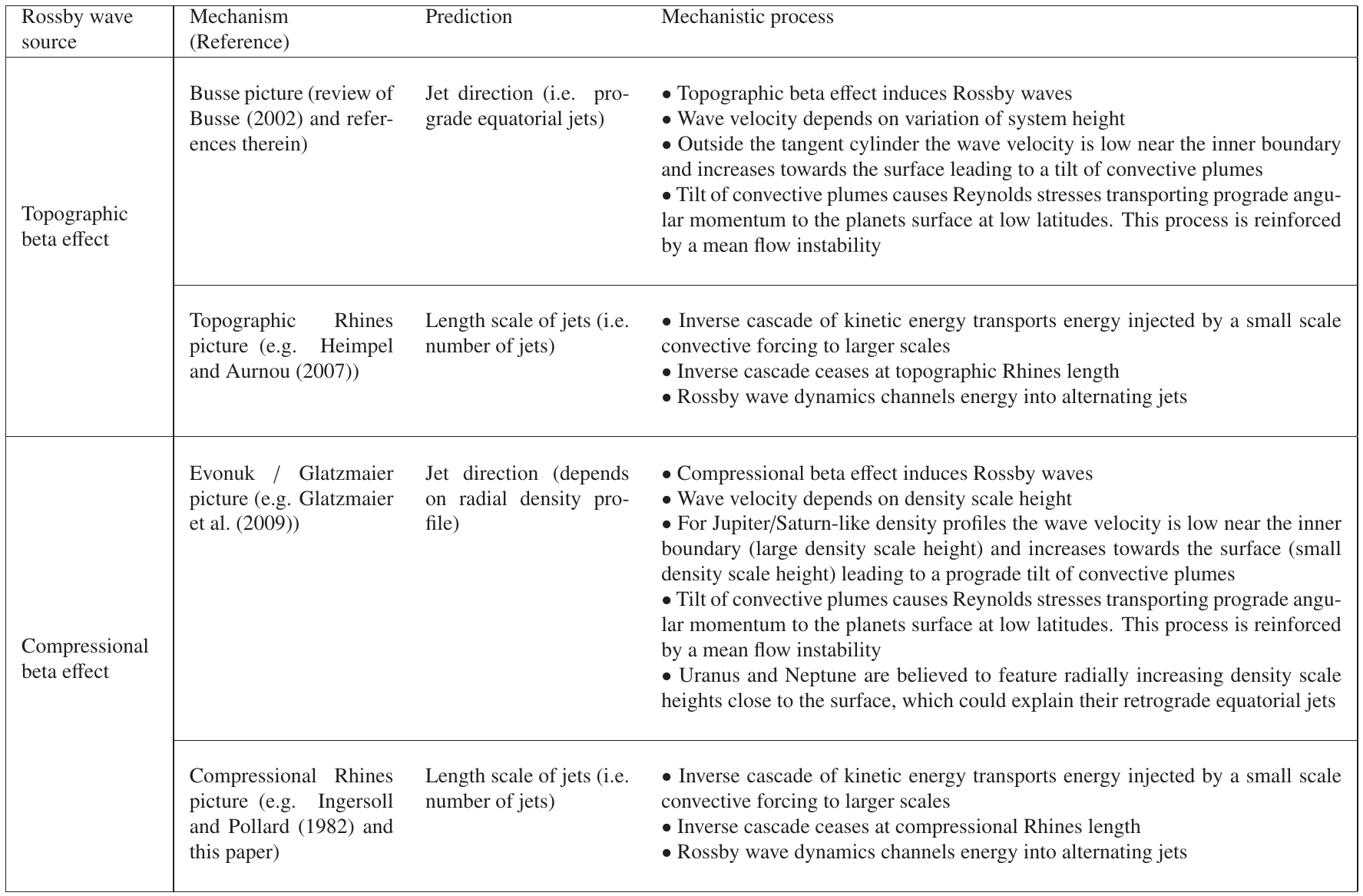}
\end{center}
\caption{List of popular mechanisms possibly driving zonal winds in giant planets in the deep-forcing scenario. The above mechanisms can be subdivided into two groups that consider different Rossby wave sources, i.e. the topographic- and the compressional beta effect. Within each group, different mechanistic pictures of the jet generation process exist, which are discussed in more detail in the text. Note that these mechanistic views may in many respects be complementary and do not necessarily oppose each other.}
\end{table*}

The vast majority of studies performed so far are based on the {\em topographic beta effect} as the source of Rossby waves. Since the large scales in rapidly rotating flows are strongly affected by Coriolis forces, coherent columnar structures parallel to the rotation axis can be expected to exist in giant planets.
Such fluid columns may extend through the entire planet touching the spherical boundaries.
When moved perpendicular to the rotation axis, they undergo vertical stretching or compression due to the spherical geometry.  This, through angular momentum conservation, results in a change of their vorticity, a process often called {\em topographic beta effect}. Since the newly generated vorticity is out of phase with the vorticity associated with the outward and inward movement of the fluid columns, azimuthally propagating Rossby waves can be generated.  

As shown by \citet{Busse1970}, convection is most easily excited outside an imaginary cylinder parallel to the rotation axis enclosing the inner spherical shell, the so-called tangent cylinder. Due to the topographic beta effect, the flow takes the form of propagating convection waves, which are essentially Rossby waves modified by the thermal buoyancy field (so-called {\em thermal Rossby waves}). Because of the curved spherical boundaries, the strength of the topographic beta effect depends on the distance from the rotation axis, which causes a slower wave propagation close to the tangent cylinder than further outside.  This in turn leads to a tilt of the convection columns in prograde direction towards the outer boundary, which gives rise to Reynolds stresses transporting prograde momentum outwards and retrograde momentum inwards. A zonal flow is generated, which is reinforced by a mean flow instability that can drive a strong prograde equatorial jet (see e.g. Busse, 2002, for a review). \citet{Busse1983,Busse1994} further argued that multiple zonal jets can be generated because the limited radial extent of the convection columns allows for a breakup of convection into cylindrical sub-layers, such that the differential rotation pattern observed on Jupiter may be explained. In table \ref{mechanism_table}, we refer to this view of differential rotation generation as the {\em Busse picture}.

Numerical simulations \citep{Heimpel2005,Heimpel2007,Jones2009} have shown that an alternating jet pattern very similar to the one observed e.g. on Jupiter, can form if the ratio of the inner to outer shell radius is chosen large enough, i.e. if the shell is assumed sufficiently thin. Typically, a strong prograde jet is found at the equator, which is readily explained by the ideas described above. The high latitude jets observed in the simulations are interpreted by the authors as being a consequence  of a Rhines-type mechanism, in which energy is transported to the large flow scales by an inverse cascade. As in the shallow models, the upscale transport finally comes to a halt when Rossby waves become important in the dynamics and channel the kinetic energy into zonal jets. The number of the jets observed in the simulations has indeed been shown to follow an approximate Rhines-type scaling. In table \ref{mechanism_table}, we call this view of zonal flow generation the {\em topographic Rhines picture}.
 
 The topographic beta effect, i.e. the local vorticity generation due to a change in system height, is not the only process possibly generating Rossby waves in deep planetary atmospheres. If the strong density stratification in giant planets is taken into account, local vorticity can also be produced in another way. As fluid parcels rise and sink in the background pressure field, they expand and contract. Because the Coriolis force exerts a torque on such fluid parcels, vorticity is locally produced or destroyed \citep{Glatzmaier1981,Glatzmaier2009}. This  local vorticity source can be interpreted as a beta-type effect \citep{Ingersoll1982}, and since it crucially relies on the compressibility of the planetary gas, we call it the {\em compressional beta effect}. The vorticity generation is again out of phase with the vorticity associated with the upwelling and downwelling fluid, such that Rossby waves can be generated. Different from the topographic beta effect however, the propagation speed of these Rossby waves is proportional to the local inverse density scale height, and is not dictated by changes of the system height. 
 
An attractive feature of the compressional beta effect is that it does not require large-scale coherent flow structures that touch the spherical boundaries in order to operate. Instead it entirely relies on the local expansion and compression of the fluid. Indeed, \citet{Glatzmaier2009} argue that the topographic beta effect is hard to maintain in the presence of the vigorous turbulence expected in planetary interiors. In their view, coherent fluid columns connecting the boundaries are likely to be shred apart by the vigorous convection, especially in planetary interiors which exhibit a strong density stratification which further facilitates the break-up of convection columns. The slope of the remote boundaries thus seems unlikely to efficiently generate vorticity in the deep interior, which, as argued by \citet{Glatzmaier2009}, casts doubt on the relevance of the {Busse-} and the {topographic Rhines picture} for highly turbulent, strongly density stratified planetary atmospheres. Models based on the compressional beta effect offer an interesting alternative here. 
 
Recently, \citet{Evonuk2006,Evonuk2007,Evonuk2008,Glatzmaier2009} and \citet{Evonuk2012} put forward a model that shares many similarities with the Busse picture, but relies on the compressional beta effect instead of the topographic one. Structure models of giant planets reveal that the density scale height in giant planets changes considerably with depth. Since the phase velocity of the Rossby waves generated by the compressional beta effect depends on the local density scale height, the waves must be expected to propagate with different phase speed at different depth, which, as in the Busse picture, would tilt convective plumes. For Jupiter- or Saturn-like density profiles, Rossby waves propagate faster near the surface than in the deep interior. Again similar to the Busse picture, the arising Reynolds stresses then transport prograde momentum outwards and retrograde momentum inwards, a process that is subsequently reinforced by a mean flow instability. This naturally explains the strong prograde equatorial jets of Jupiter and Saturn.  The retrograde jets found on Uranus and Neptune may also be explained by density profiles with radially increasing density scale heights close to the surface, as suggested by models of their interior structure \citep{Hubbard1991}.  These ideas have been tested using two-dimensional numerical simulations in equatorial planes using the anelastic approximation \citep{Evonuk2006,Evonuk2007,Evonuk2008,Glatzmaier2009,Evonuk2012}. The simulations with an inner core typically reveal the formation of two jets, one touching the planetary surface at the equator, and another one of opposite direction closer to the rotation axis, with the jet directions being controlled by the background density profile as expected. Moreover, fully convective models without an inner core allow for the occurrence of three jets. The authors of these studies provide a detailed description of the mechanistic processes generating these jets.
We refer to this view of zonal flow generation by the term {\em Evonuk / Glatzmaier picture} in table \ref{mechanism_table}. Interestingly, \citet{Evonuk2012} briefly mention that for rapid rotation and weak convective forcing, the simulations show a tendency to develop more jets, without further discussing such multiple jet states. 
 
Taking into account the similarities between the Evonuk / Glatzmaier model with the Busse picture, and the related nature of the corresponding Rossby waves, it seems likely that the topographic Rhines picture should also have a compressional equivalent. As already suggested in the early works of \citet{Ingersoll1982} and \citet{Ingersoll1986}, who mainly focussed on the stability properties of barotropic jets, such a {\em compressional Rhines mechanism} may then be expected to drive multiple zonal flows. In table \ref{mechanism_table}, we call this view of zonal flow generation the {\em compressional Rhines picture}. A detailed description will be provided in section \ref{theory}. So far, convincing experimental or numerical validation of this concept is still missing. In this paper, we aim to close this gap. 
 
The numerical evidence presented in the following sections suggests that the compressional Rhines picture provides an interesting avenue for combining the attractive features of the Evonuk / Glatzmaier view with the ability to explain multiple jets in a straight-forward manner. A number of questions immediately arise in this context. Most importantly, does a compressional Rhines mechanism indeed generate multiple jets in rapidly rotating, convective turbulence?
The answer is far from obvious, because in convective systems, the strength and spatial distribution of the turbulent forcing is intimately tied to the large scale dynamics. The natural tendency of convective turbulence to radially mix the fluid homogeneously has to be overcome in order to allow for a radially inhomogenous potential vorticity (PV) distribution, as required for jets. In particular, it appears unclear whether eddy-transport barriers, as envisioned e.g. by \citet{Dritschel2008}, can be maintained against incursions of vigorous convective plumes. If jets can indeed be generated in the proposed way: How well does the usual picture of turbulence-wave crossover describe the dynamics in direct numerical simulations? Do the jets generated by a compressional Rhines mechanism follow a  compressional Rhines scaling? What are the numerical parameters characterizing such a scaling law?  Do predictions based on this scaling broadly fit the planetary jets observed in our solar system? Preliminary answers to these question, based on a highly idealized \jan{2-d} model of planetary convection, will be given in the following sections. 

 Our paper is organized as follows. To give the reader a visual idea of the physical effects discussed in this work, we first describe some observations from numerical simulations of compressible (anelastic) rotating convective turbulence developing into multiple jets (section \ref{observation}). A more formal definition of the model considered in this paper is given in section \ref{method}, which is then used to develop a simple theoretical model for the observed jets in section \ref{theory}. The theoretical predictions are compared with the results from a large number of numerical simulations in section \ref{results}. Following some speculations on the implications of our work for planetary interiors in section \ref{speculations}, general conclusions are given in section \ref{conclusions}.

\section{Observation}
\label{observation}

\begin{figure}
\includegraphics[width=\linewidth]{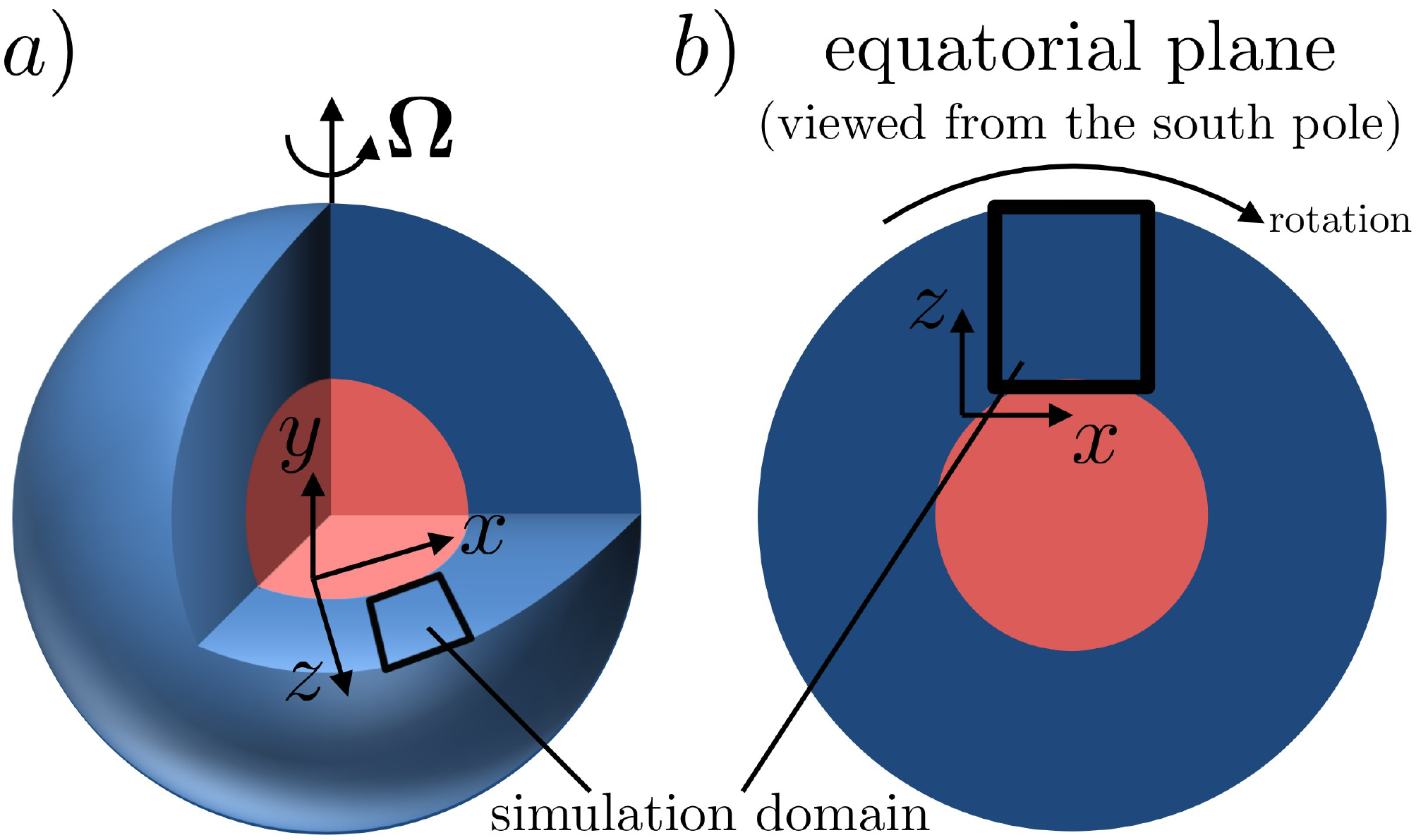}
\caption{As a crude model for the convective flow in the equatorial plane of a rapidly rotating giant planet (a), we consider two-dimensional flow in a small equatorial section\alt{(b)}, which is approximated as a plane layer \neu{(b)}. A fixed entropy contrast between \neu{the planet's inner part (red/light)}\alt{an inner core} and the surface drives thermal convection \neu{within the simulation domain}. In the plane layer, horizontal mean flows in $x$-direction represent eastward, i.e. prograde, winds, and correspondingly, flows in the opposite direction represent westward, i.e. retrograde jets. \neu{The radial direction coincides with the $z$-direction in the Cartesian cutout on the equatorial plane and the rotation vector $\vec\Omega$ is parallel to the $y$-axis.} The anelastic approximation is used to retain the dynamical effects of the compressibility of the planetary gas. (Color version available in the online version of Icarus.)}
\label{setup}
\end{figure}

To give the reader a visual idea of the physical effects discussed in this paper, we start with describing some key observations from numerical simulations. It is important to make clear from the outset that we are not attempting to simulate convection in giant planets as realistically as possible. Instead, our goal here is to devise a simple model that illustrates the dynamical role of compressibility in driving zonal winds with maximum clarity. We crudely model convection in an  equatorial plane of a rapidly rotating giant planet by considering a two-dimensional, Cartesian cutout, as shown in figure \ref{setup}. Periodic boundary conditions are assumed in the horizontal direction.  For simplicity, we neglect the curvature of the spherical boundaries and the variation of gravity with depth. The planetary gas is assumed to be heated from below and cooled from above. Since the flow speed is typically much smaller than the sound speed, the anelastic approximation is employed. The assumption of two-dimensionality is motivated by the rapid rotation of the planet, which is expected to damp flow variations along the rotation axis.

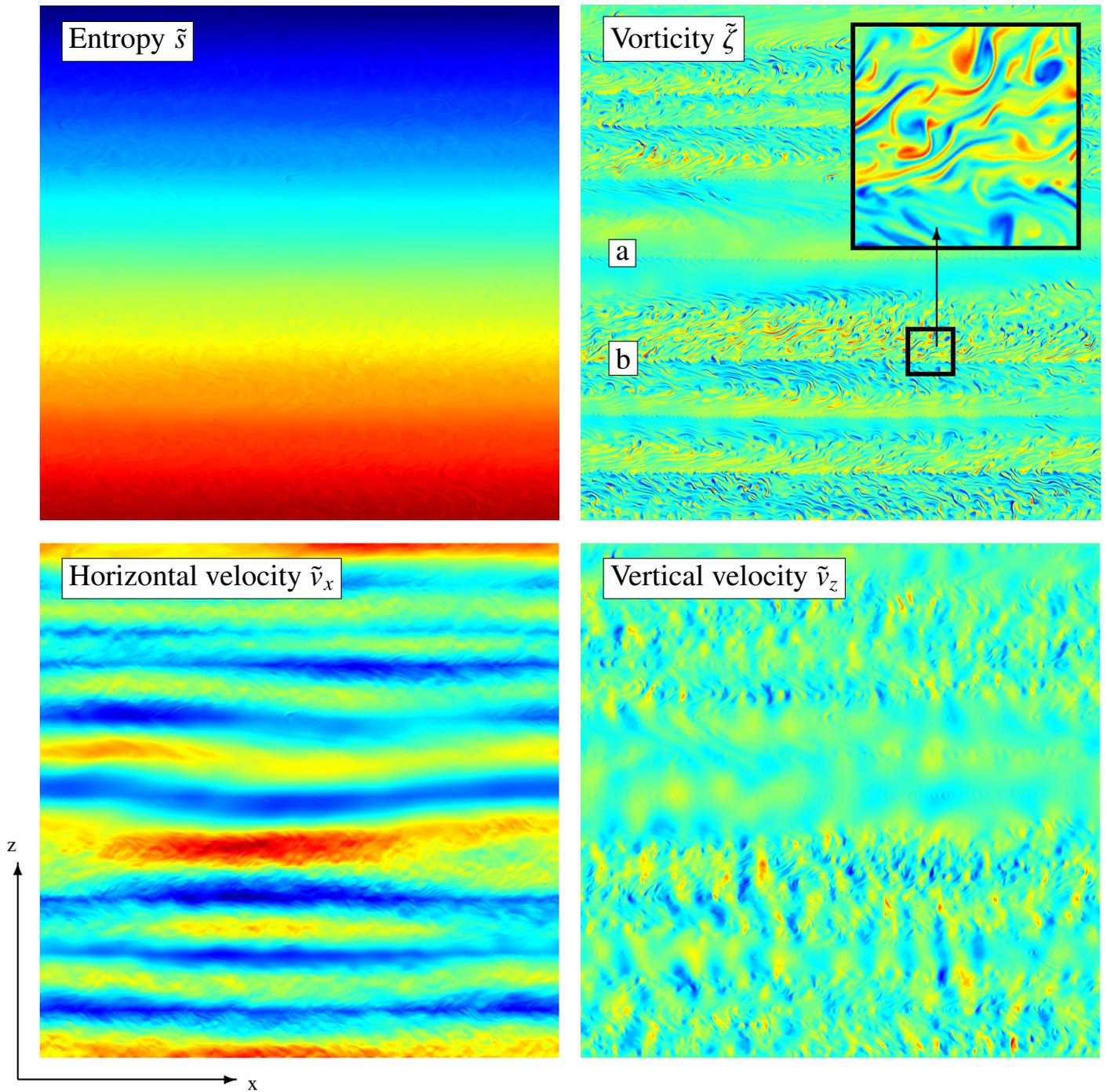
\begin{figure*}
\setlength{\unitlength}{\linewidth}%
\begin{picture}(1.0,1.0)
\thicklines
\put(0.03,0.525){\includegraphics[width=0.475\linewidth]{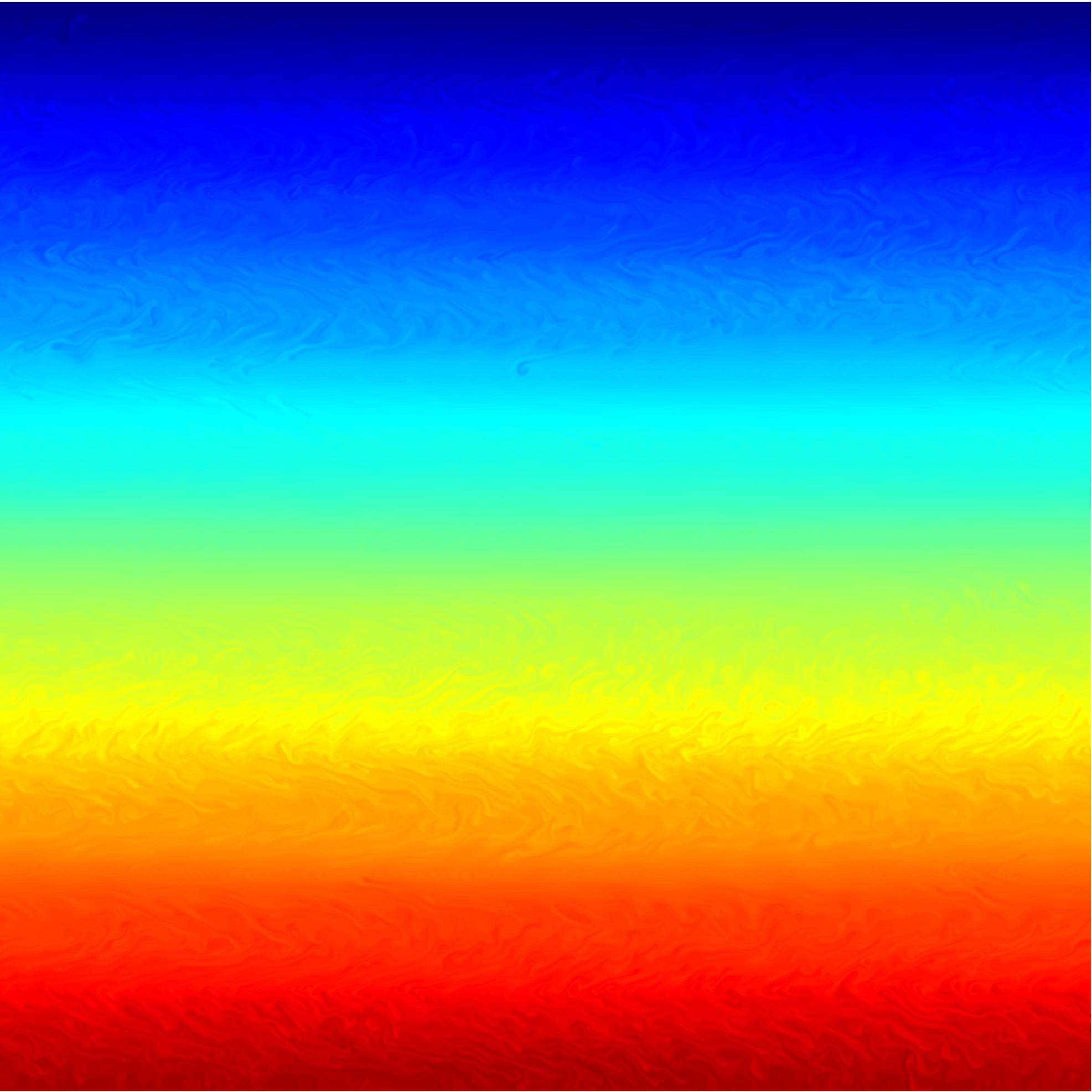}}
\put(0.525,0.525){\includegraphics[width=0.475\linewidth]{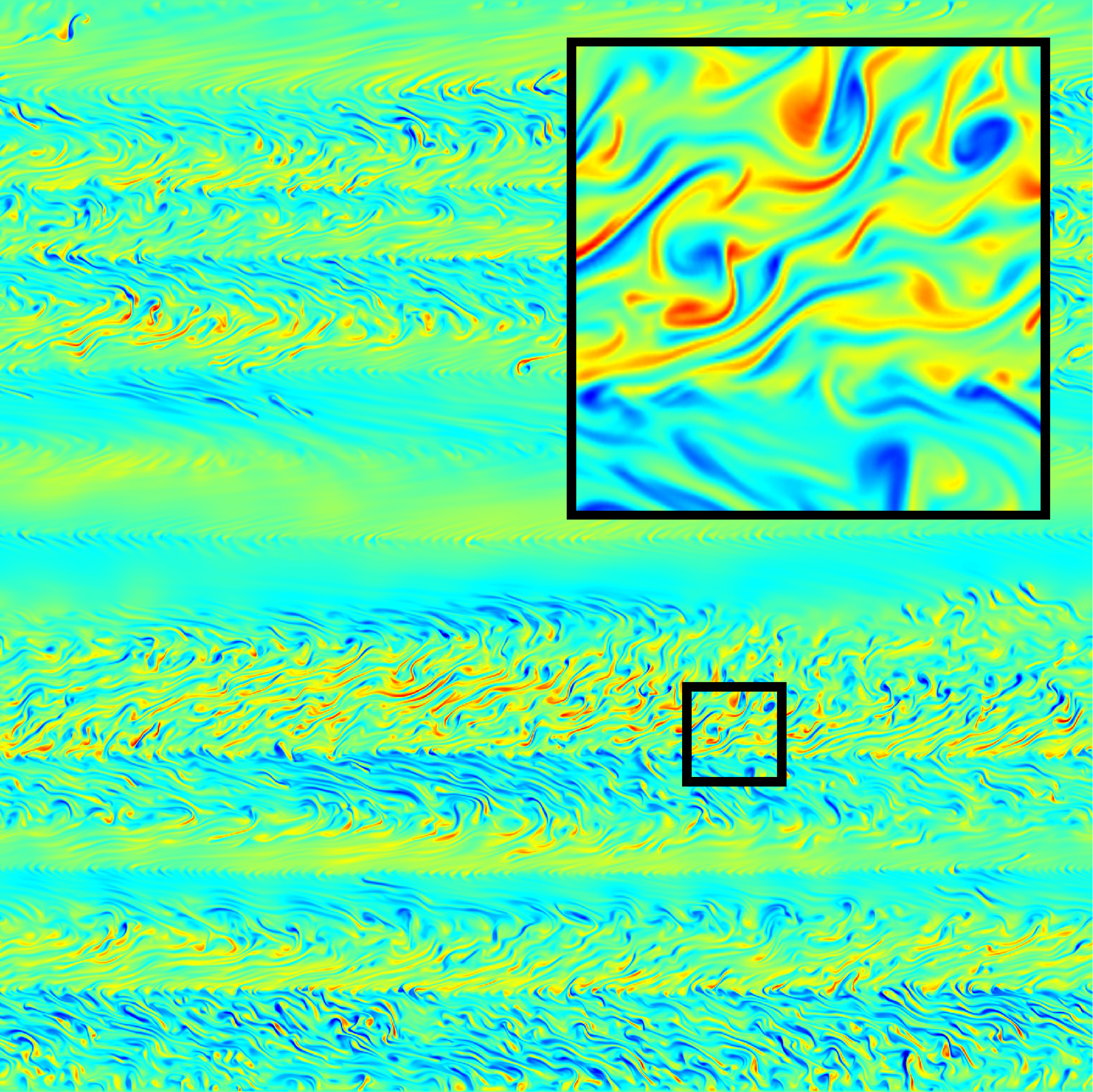}}
\put(0.03,0.03){\includegraphics[width=0.475\linewidth]{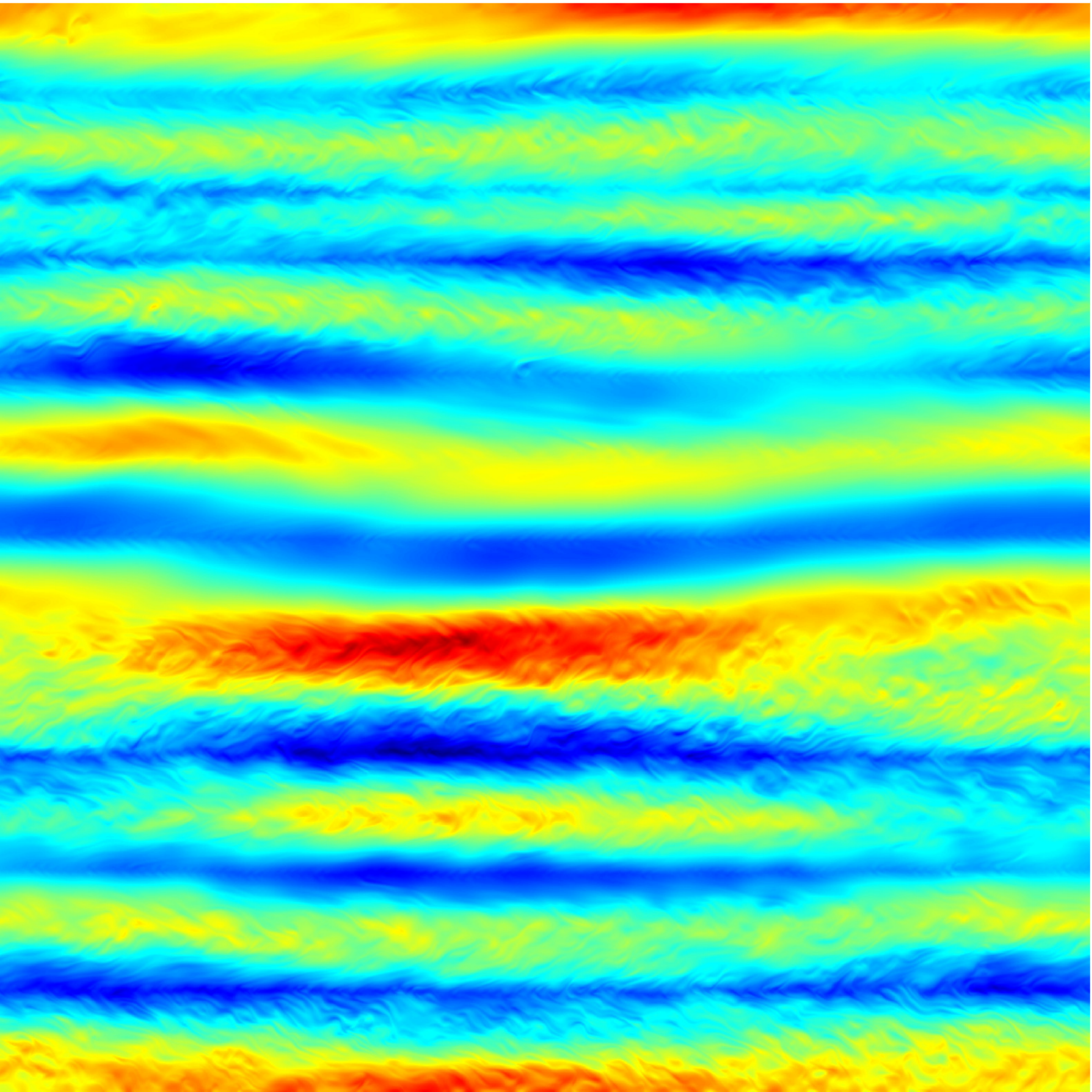}}
\put(0.525,0.03){\includegraphics[width=0.475\linewidth]{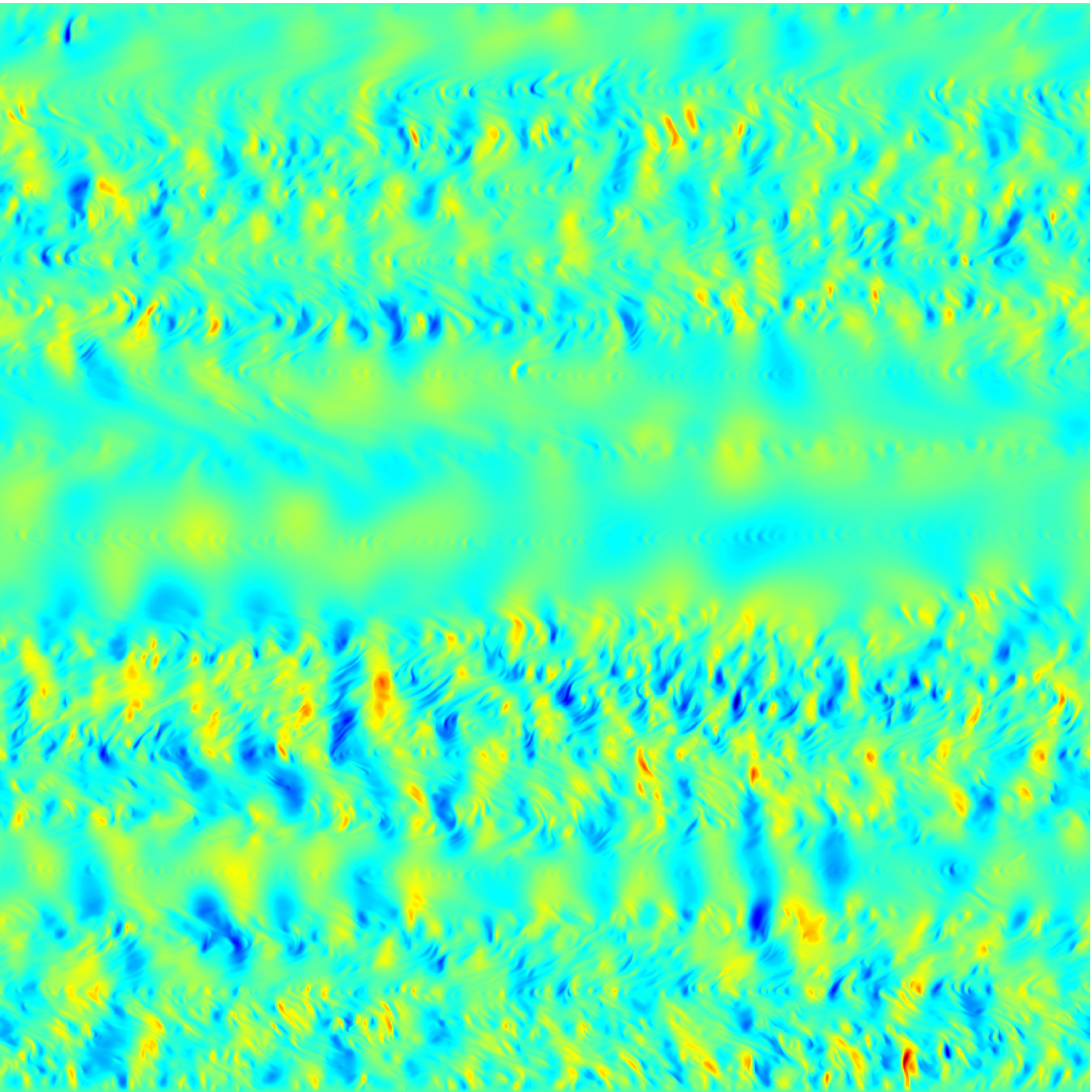}}
\put(0.55,0.765){\fcolorbox{black}{white}{\Large a}}
\put(0.55,0.665){\fcolorbox{black}{white}{\Large b}}
\put(0.05,0.96){\fcolorbox{black}{white}{\Large Entropy $\tilde{s}$}}
\put(0.545,0.96){\fcolorbox{black}{white}{\Large Vorticity $\tilde{\zeta}$}}
\put(0.05,0.465){\fcolorbox{black}{white}{\Large Horizontal velocity $\tilde{v}_x$}}
\put(0.545,0.465){\fcolorbox{black}{white}{\Large Vertical velocity $\tilde{v}_z$}}
\put(0.85,0.685){\vector(0,1){0.11}}
\put(0.01,0.01){\vector(1,0){0.2}}
\put(0.01,0.01){\vector(0,1){0.2}}
\put(0.22,0.00){x}
\put(0.00,0.22){z}
\end{picture}
\caption{Snapshots of the entropy and vorticity field and of the velocity components for a typical simulation. The entropy field represents the buoyancy of fluid parcels, with red colors denoting warm, buoyant material while blue signifies cold fluid. In the remaining panels, large positive values are denoted by dark red, and large negative values by dark blue, with green colors depicting values close to zero. In contrast to simulations using incompressible fluids, strong horizontal jets dominate the dynamics. The parameters corresponding to the snapshots, defined in section \ref{method}, are \neu{$Pr=1$,} $Ek = 2.5\cdot 10^{-10}$, $Ra = 3\cdot 10^{12}$ and $\chi=1.2$. (Color version available in the online version of Icarus.)}
\label{snapshots}
\end{figure*}

We start the simulations from a purely conductive state, which quickly becomes unstable and is replaced by vigorous, turbulent convection. The rapid rotation confines the fluid flow to relatively small spatial scales and omits up- and downwellings from crossing the entire layer. Soon after convection becomes fully established, multiple alternating jets spontaneously emerge in the flow field. Drawing energy from the small scale convective motions, they grow until they become the most dominant feature of the flow. Snapshots from a typical simulation (parameters are discussed in section \ref{method}) are shown in figure \ref{snapshots}. About 15 individual jets, superimposed by smaller-scale convective motions, are clearly visible. 

\begin{figure*}
\includegraphics[width=\linewidth]{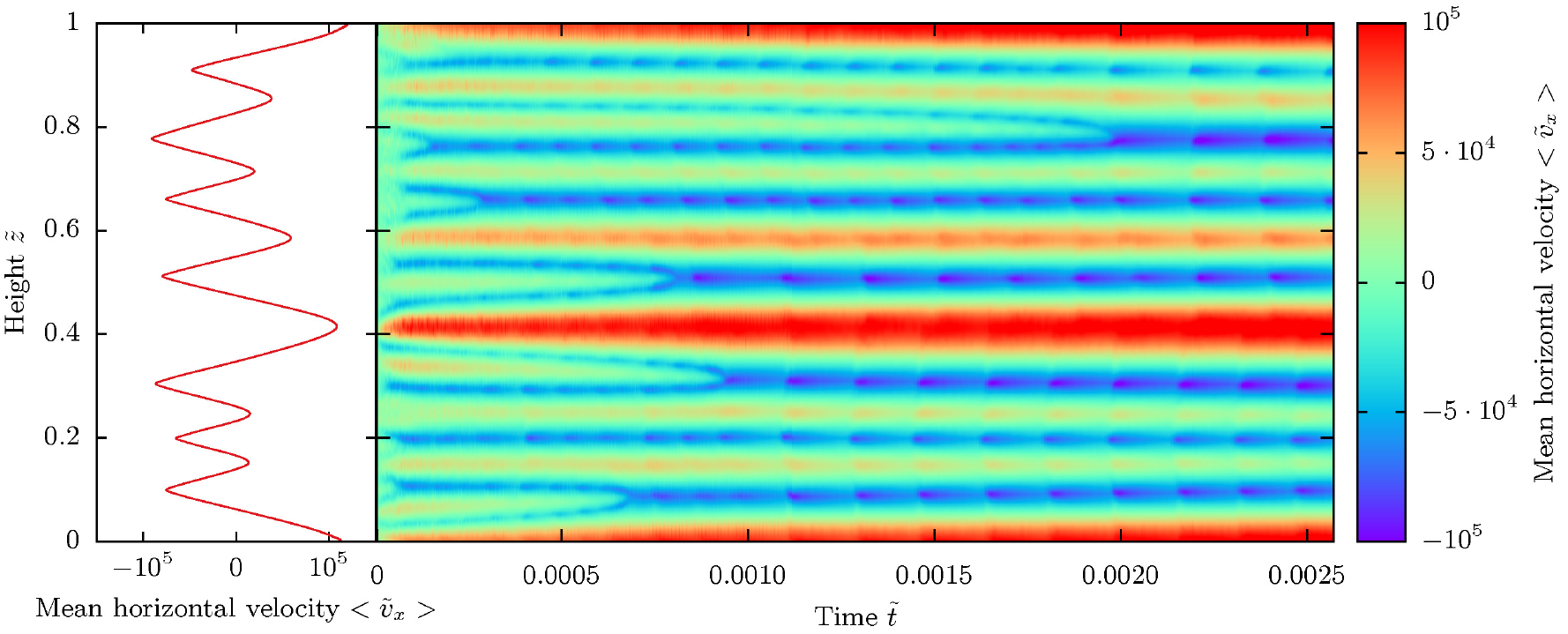}
\caption{The horizontal mean velocity (jet speed) is plotted against height (planetary radius) for the same model simulation as shown in figure \ref{snapshots}. The left plot displays the time averaged mean velocity profile from the time span after the last merging event, while the plot on the right shows the temporal evolution of the jets. The simulation has been carried out over a time span much longer than the time scale on which the first jets form. Apart from sporadic merging events, the jets are stable over the time span covered by the simulation. Non-dimensional units, defined in detail in sec. \ref{sec:non_dimensionalization}, are used. (Color version available in the online version of Icarus.)}
\label{mean_vx}
\end{figure*}

Figure \ref{mean_vx} illustrates the temporal evolution of the jets for the same simulation, and also shows a vertical  profile of the mean horizontal velocity at an advanced stage. After the initial jet formation, the vertical position of the jet maxima remains nearly constant, interrupted only by sporadic merging events. A closer inspection reveals a\alt{ noticeable} pulsation of the jet velocities, which is caused by an interaction with the underlying convective flow. As the jet amplitude grows, the associated shear successively destroys the convective eddies, which in turn deprives the jet from its driving. The jet then decays on a viscous time scale, until convective instabilities again begin to grow exponentially, resulting in a convective burst and a subsequent acceleration of the mean flow. This can also be seen in figure \ref{snapshots}, where convection is almost destroyed by the shear in the region denoted by (a), whereas a convective burst, accompanied by strong vortical eddies, occurs in region (b). It also becomes evident in time series of the heat flow and of the kinetic energy of the jets, as shown in figure \ref{nusselt_meanenergy}, which will be discussed in more detail in section \ref{results}. Such relaxation oscillations are typical when convection generated shear interacts with the \neu{radial}\alt{vertical} flow, and have been observed in a variety of situations \neu{(e.g. \citet{Brummell1993,Aurnou2001,Christensen2001b,Grote2001,Christensen2002,Busse2002,Jones2003,Simitev2003,Morin2004,Rotvig2006,Ballot2007,Gastine2012,Heimpel2012,Teed2012})}.

The results shown in figures \ref{snapshots} and \ref{mean_vx} clearly demonstrate that multiple jets can be driven in our simple model, an effect which is not observed in  corresponding Boussinesq cases. The compressibility of the gas thus promotes jet formation and allows a large number of jets to be generated. 
Further simulations reveal that the number of jets is controlled by the rotation rate, the amount of density increase with depth and by the strength of the convective forcing. A slower rotation, less variation in density and an increase of the convective vigor tend to result in fewer jets. The remainder of this paper is devoted to analyzing the effects described in this section.

\section{Model}
\label{method}

We begin by defining our simple model, which was illustrated in figure \ref{setup}, more formally. Convection in a two-dimensional Cartesian layer, rotating about an axis perpendicular to the layer plane, is studied. For simplicity, gravity is assumed to be constant, and the fluid is assumed to be a polytropic ideal gas. To account for the vertical compression of the gas without sacrificing numerical efficiency, we employ the anelastic approximation. This has the advantage that fast acoustic modes, which are believed to be unimportant for planetary convection, are filtered out from the outset. Note that several versions of the anelastic approximation, varying in complexity, have been developed  \citep{ogura1962,Gough1969,gilman1981compressible,Braginsky1995,Lantz1999}. Here, we use the version proposed by \citet{Lantz1999}. This choice is purely motivated by the simplicity of the resulting equations, which allows for an efficient numerical treatment. \neu{We stress again that our model is designed to isolate the dynamical effects driving zonal winds induced by the compressibility of the gas atmosphere, and not to study convection in planetary interiors in a realistic setting (implying a spherical geometry and other effects possibly driving jets)}\alt{We stress again that our model is designed to isolate the dynamical effects driving zonal winds in a vertically compressed gas atmosphere, and not to study convection in planetary interiors in a realistic setting}.

\subsection{Governing equations}
\label{governing_equations}

The thermodynamic quantities (density $\rho$, pressure $p$, temperature $T$ and entropy $s$) are decomposed into a static reference state (subscript $0$) and a small time-dependent perturbation (subscript $1$)

\begin{align}
\nonumber
\rho=\rho_0(z)+\rho_1(t,x,z), \qquad & p=p_0(z)+p_1(t,x,z),\\
T=T_0(z)+T_1(t,x,z), \qquad & s=s_0(z)+s_1(t,x,z).
\end{align}
The reference state is assumed to be constant in time, while the perturbations are governed by the equations
\begin{equation}
\nabla\cdot\left(\rho_0 \vec v\right) = 0,
\label{continuity}
\end{equation}
\begin{align}
\nonumber
\frac{\partial\vec v}{\partial t}+\left(\vec v\cdot\nabla\right)\vec v = & - \nabla\left(\frac{p_1}{\rho_0}\right) + \frac{g}{c_p} ~ s_1 ~ \vec e_z \\ & + \frac{1}{\rho_0}\left(\nabla \cdot \Pi\right) - 2\Omega ~ \vec e_{y}\times\vec v,
\label{momentum}
\end{align}
\begin{equation}
\rho_0 T_0 \left(\frac{\partial s_1}{\partial t}+\left(\vec v\cdot\nabla\right) s_1\right) = \nabla\cdot\left(\kappa_0\rho_0 T_0\nabla s_1\right) + \left(\Pi \cdot\nabla\right)\cdot\vec v,
\label{heat}
\end{equation}
which express mass, momentum and energy conservation, respectively. Here, $t$ denotes time, $\vec v$ is the velocity and  $\kappa_0(z)$ is the background turbulent eddy diffusivity of entropy. The tensor 
\begin{equation}
\Pi_{ij}=\mu_0(z)\left(\frac{\partial v_i}{\partial x_j}+\frac{\partial v_j}{\partial x_i}-\frac{2}{3}\nabla\cdot\vec v\delta_{ij}\right)
\end{equation}
is the Newtonian viscous stress tensor with $\mu_0(z)$ denoting dynamic viscosity. The acceleration of gravity $g$, the specific heat at constant pressure $c_p$ and the angular velocity vector $\Omega ~ \vec e_y$ are assumed to be constant in time and space. $\vec e_y$ and $\vec e_z$ are the unit vectors.

\subsection{Isentropic background state}
As conventionally assumed, the thermodynamic background state is taken to be isentropic. By applying the first law of thermodynamics in the form $\rho T ds = \rho c_p dT - \delta_p dp$,  with $\delta_p=-(\partial \ln \rho /  \partial \ln T)_{p}$ being the dimensionless thermal expansion coefficient ($\delta_{p}=1$ for an ideal gas), using the hydrostatic condition $dp_0/dz=-\rho_0 g$ and the ideal gas law $p=(c_p-c_v)\rho T$, we find 
\begin{align}
\label{back_temp}
T_0(z)= & T_r\left(1-\frac{g}{c_p T_r}z\right), \\
\label{back_dens}
\rho_0(z)= & \rho_r\left(1-\frac{g}{c_p T_r}z\right)^n,
\end{align}
where the index $r$ denotes reference values defined here at the bottom boundary and $c_v$ is the specific heat at constant volume. Following the usual notation, $n=c_v/(c_p-c_v)$ is the polytropic index. Note that at this point, the vertical variation of dynamic viscosity $\mu_0$ and turbulent entropy diffusivity $\kappa_0$ can still be chosen freely. 

Most anelastic simulations of planetary convection published so far \citep{Evonuk2004,Glatzmaier2009,Jones2009,Kaspi2009,Gastine2012} have assumed constant values for kinematic viscosity $\nu_{0}=\mu_{0}/\rho_{0}$ and for entropy diffusivity $\kappa_{0}$. For the sake of comparability,  we follow this approach here. Another obvious choice would have been to use a constant dynamic viscosity while assuming an entropy diffusion that is inversely proportional to the density background \citep{Lantz1999,rogers2003simulations}. We do not expect the precise choice to have a significant impact on the nature of the dynamical effects studied here.

\subsection{Boundary conditions}
A fixed entropy contrast $\Delta s$ is maintained between the lower and upper boundaries, which are further assumed to be impermeable and shear stress free. To minimize the influence of the horizontal boundaries, periodicity is assumed in the horizontal direction. 

\subsection{Non-dimensionalization}
\label{sec:non_dimensionalization}
In order to solve the governing equations (\ref{continuity}-\ref{heat}) numerically, they are non-dimensionalized using the layer height $d$ as the length scale, the thermal diffusion time $d^2/\kappa_r$ as the time scale and the constant entropy difference $\Delta s$ between the top and bottom boundary to scale entropy. The system is then characterized by the five dimensionless control parameters
\begin{align}
\nonumber
Ra & =\frac{g\rho_r \Delta s d^3}{c_p \mu_r \kappa_r}, & Pr & =\frac{\mu_r}{\rho_r \kappa_r}, & Ek =\frac{\mu_r}{2\Omega d^2 \rho_r}, \\
\chi & =\frac{\rho_{bot}}{\rho_{top}}=(1-D)^{-n}, & n  & =\frac{c_v}{c_p-c_v}, 
\end{align}
where $Ra$ is the Rayleigh number measuring the strength of buoyancy forcing, $Pr$ is the Prandtl number relating the viscous to the thermal diffusivity and  $Ek$ is the Ekman number comparing the rotational to the viscous timescale. $\chi$ denotes the density contrast between the bottom and top boundary. Note that for an ideal gas, it can be related to the Dissipation number $D=g d/(c_p T_r)$. As before,  $n$ denotes the polytropic index. Variables using this non-dimensionalization will be marked with a tilde, e.g.  $\tilde{v}$, in the following. 
\subsection{Numerical technique}
The simulations presented in this paper have been performed 
with an anelastic code which is a modified version of the Boussinesq code by \citet{Stellmach2008}. It uses a mixed pseudo-spectral 4th-order-finite-difference spatial discretization and a semi-implicit time stepping scheme based on a third order Adams Bashforth/backward-difference formula (AB3/BDF3). All linear terms including the Coriolis force are treated implicitly.

\section{Compressional beta effect}
\label{theory}

In this section the vorticity equation governing the dynamics of compressible (anelastic) turbulent convection is discussed. We will show that, under certain assumptions, it can be manipulated into a form which is mathematically equivalent to the vorticity equation encountered in two-dimensional beta plane turbulence. A detailed discussion of the assumptions made will help to constrain the situations in which multiple jets are to be expected. 

\subsection{Anelastic vorticity dynamics}
\label{vorticity_dynamics}
Taking the curl of equation (\ref{momentum}) and invoking  equation (\ref{continuity}), we find  
\begin{align}
\nonumber
{\frac{\partial \zeta}{\partial t}} = & - {\left(\vec v \cdot \nabla\right) \zeta} - {\frac{g}{c_p} \frac{\partial s}{\partial x}} +{\nabla \times \left(\frac{1}{\rho_0}\nabla\cdot \Pi\right)} \cdot \vec e_y \\
& + {2\Omega \rho_0^{-1}\frac{d\rho_0}{dz} v_z} + {\rho_0^{-1}\frac{d\rho_0}{dz}\zeta v_z},
\label{vorticity}
\end{align}
where $\zeta$ denotes the y-component of vorticity. The terms on the right hand side describe, from left to right, vorticity advection, vorticity generation by buoyancy forces, viscous effects, a term describing vorticity production by the Coriolis force acting on compressing and expanding fluid parcels and a corresponding nonlinear term. Focussing on a particular length scale $l_s$, we scale equation (\ref{vorticity}) by using velocity, time and entropy scales $v_s,t_s, s_s$ typical for features on that length scale and corresponding scales $\mu_s$ for viscosity and $\rho_s$ for density, resulting in 
\begin{align}
\nonumber
\frac{l_s}{t_s v_s}\frac{\partial \zeta^*}{\partial t^*} = & - \left(\vec v^* \cdot \nabla^*\right) \zeta^* + B \frac{\partial s^*}{\partial x^*} \\
\nonumber
& + \frac{1}{Re}\nabla^* \times \left(\frac{1}{\rho_0^*}\nabla^*\cdot \Pi^*\right) \\
& - \frac{1}{Ro} l_sH_\rho^{-1} v_z^* - l_sH_\rho^{-1}\zeta^* v_z^*,
\label{vorticity2}
\end{align}
where a star denotes non-dimensional O(1) quantities.
 Here,  $H_\rho=-\left((d\rho_0/dz)/\rho_0\right)^{-1}$ is the density scale height, $Re=(\rho_s v_s l_s)/\mu_s$ denotes the scale dependent Reynolds number, $Ro=v_s/(2\Omega l_s)$ is the scale dependent Rossby number and $B=-g l_s s_s/(v_s^2 c_p)$ is the buoyancy number.

The Reynolds number will typically be large except for features on the tiny dissipation scale, where viscosity becomes important. On larger scales, viscosity can be neglected. The nonlinear vorticity production term may be neglected for  $l_s H_\rho^{-1} \ll 1$, i.e. on scales $l_s$ much smaller than the density scale height. However, the linear Coriolis term can still remain important on such scales as long as the local Rossby number $Ro$ remains at least as small as $l_s H_\rho^{-1}$. 

The jets forming in the numerical simulation discussed in section \ref{observation} indeed have width which are much smaller than $H_\rho$, and they are powered by even smaller convective eddies.  It thus seems reasonable to focus on the case $l_s H_\rho^{-1} \ll 1$ in order to investigate the dynamics observed in the simulation. Neglecting viscosity and nonlinear Coriolis effects, the vorticity equation can then be written in the form 
\begin{equation}
\frac{\partial \zeta}{\partial t} + \left(\vec v \cdot \nabla\right) \zeta + \beta_\rho v_z = F,
\label{vorticity3}
\end{equation}
with $ F \equiv -({g} / {c_p}) {\partial s}/{\partial x} \;$ and 
\begin{equation}
\beta_\rho \equiv - 2\Omega \rho_0^{-1}\frac{d\rho_0}{dz} = 2\Omega ~ H_\rho^{-1}.
\label{beta_rho}
\end{equation}
Equation (\ref{vorticity3}) has the same mathematical form as the vorticity equation solved in 2-d incompressible beta-plane turbulence  (e.g. \citet{Rhines1975,Williams1978,Vallis1993,Chekhlov1996,Danilov2004,Sukoriansky2007}). However, while the beta-term in these cases is generated by a latitudinal variation of the tangential component of the Coriolis force, in our case it is a consequence of the background density stratification.

In order to further illustrate the mathematical analogy of both cases, we also consider mass conservation. By scaling the continuity equation (\ref{continuity}) in the same fashion as the vorticity equation (\ref{vorticity2}), we find
\begin{equation}
\nabla^*\cdot \vec v^* = l_s H_\rho^{-1} v_z^*.
\label{continuity_scaled}
\end{equation}
\alt{where $v_z^*$  is an $O(1)$ quantity. }The amount of compression or expansion of fluid particles thus depends on the vertical distance they travel. The vertical density variation causes significant volume changes only on scales exceeding $O(H_\rho)$, while much smaller eddies may be modeled as incompressible. Assuming again that $l_s H_\rho^{-1} \ll 1$ holds for the dynamically active scales in the simulation described above, we conclude that mass conservation is governed by  
\begin{equation}
\nabla\cdot \vec v = 0
\label{continuity_incompressible}
\end{equation}
to leading order, as in incompressible beta-plane turbulence. 

Finally, we note that the variation of $\beta_\rho$ as a function of $z$ is very small in the simulation. In order to keep our theoretical model as simple as possible, we therefore also replace $\beta_\rho$ by its \neu{domain}\alt{volume} average $\langle ... \rangle$, 
\begin{equation}
\langle \beta_\rho\rangle=2\Omega \left \langle H_\rho^{-1} \right \rangle.
\label{beta_vol_avrg}
\end{equation}
We will discuss the accuracy of this assumption in detail in section \ref{errors}. For the simulation in section \ref{observation}, the maximum error is smaller than about $4\%$.  Note that by neglecting the height dependence of $\beta_\rho$, we explicitly exclude the term that \neu{determines the wind directions in the Evonuk / Glatzmaier picture}\alt{drives zonal flow in the Evonuk / Glatzmaier mechanism} from our considerations\footnote{We have verified that a simulation identical to the one presented in section \ref{observation}, except for the fact that a hypothetical density stratification with constant $H_\rho$ and thus $\beta_\rho$ was employed, yields similar jet formation behavior.
\neu{Therefore, a spatially varying $\beta_\rho$ is not essential for jet generation.}\alt{ Therefore, the Evonuk / Glatzmaier mechanism may safely be excluded from the theory in the present context.}}.

Replacing $\beta_\rho$ by its \neu{domain}\alt{volume} average and introducing the stream function $\psi$ , defined by $v_x=-\partial \psi / \partial z$ and $v_z = \partial \psi / \partial x$, so that $\zeta=-\nabla^2 \psi$, we find from (\ref{vorticity3}) and (\ref{continuity_incompressible}) that 
\begin{equation}
\frac{\partial}{\partial t}\nabla^2\psi + J\left(\psi,\nabla^2\psi\right) - \langle \beta_\rho \rangle \frac{\partial}{\partial x} \psi = -F,
\label{wave}
\end{equation}
where $J(a,b)= \partial a/ \partial x ~ \partial b / \partial z - \partial a/ \partial z ~ \partial b / \partial x$ denotes the standard Jacobian. This equation is identical to the one commonly used in simple 2d incompressible shallow forcing models.
\neu{Similar to convectively driven simulations with the topographic beta effect (e.g. \citet{Brummell1993,Jones2003,Morin2004,Morin2006,Gillet2006,Rotvig2006,Rotvig2007,Teed2012})}\alt{Note however that} the forcing $F$ is not prescribed in our case, but is a function of the entropy field governed by  equation (\ref{heat}), thus giving rise to an additional non-linearity in our system. In particular, the strength of the buoyancy forcing as a function of the spatial scales \stephan{and its spatial distribution} is not \neu{known a priori}\alt{a-priorly known}. \neu{Note that we abstain from implementing the effect of Ekman pumping and suction which would result in a scale-independent damping term in equation (\ref{wave}). Such a so-called bottom friction term
has been shown to favor the occurrence of multiple jets (see e.g. \citet{Jones2003,Teed2012}). Introducing this effect may be justified for convection inside the tangent cylinder where the the inner boundary is set by the transition to metallic hydrogen and can be argued to be approximated by a rigid boundary (e.g. \citet{Jones2003}). This is more difficult to justify in our simulations which focus on the dynamics outside the tangent cylinder. Furthermore, 
the application of bottom friction is not consistent with the view that the high level of turbulence prevents the flow in the planetary interior from being strongly affected by distant boundaries  \citep{Glatzmaier2009}.  As the remote boundaries are not relevant for the vorticity generation due to the compressional beta effect, either, we keep the model simple by not invoking the possible damping effect of the boundaries.}

\subsection{Turbulence-wave crossover and \jan{compressional Rhines scale}}

Given that apart from the physical interpretation of $\beta$ and apart from the nature of the forcing, equation (\ref{wave}) is identical to the one used in many shallow forcing models, the results obtained there (see e.g. \citet{Vallis1993,Rhines1994,Vallis2006,Scott2012a,Scott2012b})  should carry over to the case studied here \neu{as experiences from simulations with convective forcing and topographic beta effect (e.g. \citet{Jones2003,Rotvig2006,Teed2012}) suggest}. In particular, neglecting the forcing term and the vertical boundaries, the equations support Rossby-like waves. This can directly be verified by looking for plane wave solutions
$\psi(t,x,z)=\psi_0 \text{exp}\left[i\left(k_x x + k_z z - \omega t\right)\right]$, yielding the dispersion relation
\begin{equation}
\omega=\frac{\langle\beta_\rho\rangle k_x}{k_x^2 + k_z^2}=\frac{\langle\beta_\rho\rangle}{k}\cos \alpha,
\label{dispersion}
\end{equation}
where $\alpha=\text{atan}(k_z/k_x)$ denotes the angle between $k_x$ and $k_z$ and $k=\sqrt{k_x^2+k_z^2}$ is the absolute wavenumber.

\begin{figure}
\setlength{\unitlength}{\linewidth}
\begin{picture}(1.0,1.0)
\put(0.0,0.0){\includegraphics[width=\linewidth]{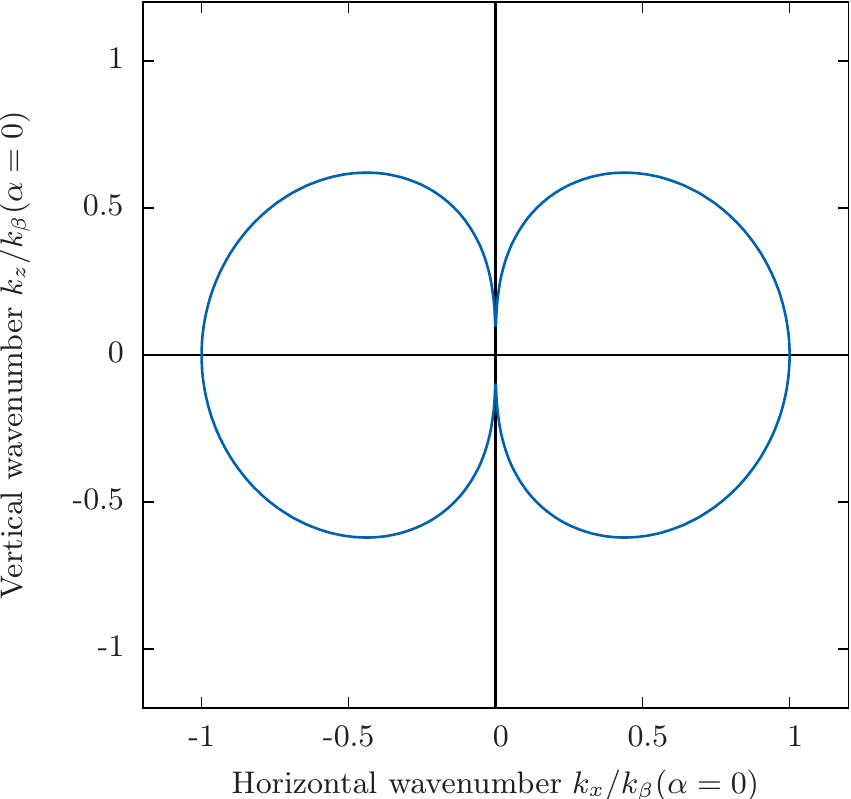}}
\put(0.65,0.85){\large \textcolor{red}{Turbulence}}
\put(0.3,0.55){\large \textcolor{red}{Rossby waves}}
\put(0.23,0.68){\fcolorbox{white}{white}{\large \textcolor{blue}{crossover}}}
\end{picture}
\caption{The theoretical Rossby-wave-turbulence crossover in wave number space is characterized by a dumbbell shape \neu{(blue)}. Two-dimensional turbulence is assumed to dominate outside the dumbbell, whereas the $\beta_\rho$-term \neu{supporting Rossby waves} is important inside. \neu{The wavenumbers are normalized by $k_\beta(\alpha=0)=\sqrt{\langle\beta_\rho\rangle / v_\beta}$, which is identical to the inverse compressional Rhines scale.}\alt{ The maximum amplitude is normalized $1$.}}
\label{dumbbell_theo}
\end{figure}
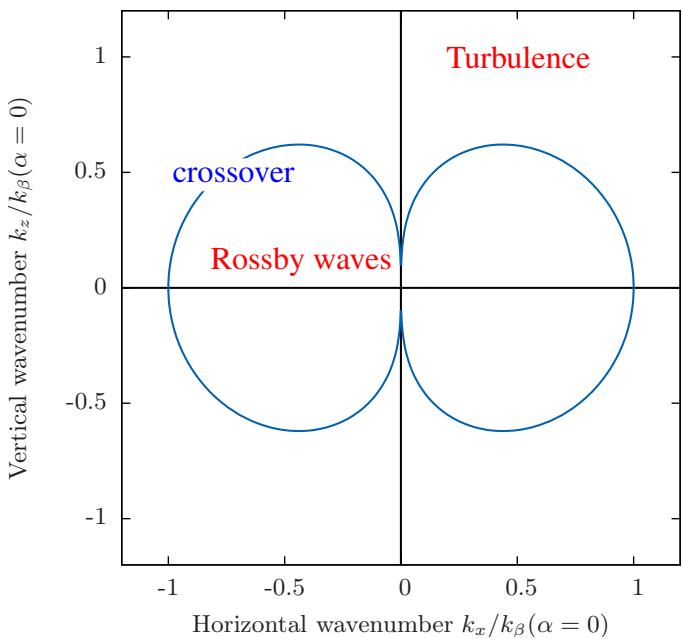

From equation (\ref{vorticity2}), it is clear that the beta-term becomes important for large scales only, more precisely for scales characterized by a local Rossby number $Ro \lesssim O(l_s H_\rho^{-1}) \ll 1$. In the simulations presented in section \ref{observation}, the first convective instabilities typically occur on much smaller scales. The dynamics on intermediate scales is then characterized by an inverse cascade \citep{Kraichnan1967,Rutgers1998}, successively transporting kinetic energy from the small forcing scale towards larger flow structures\footnote{Note that as energy is transported towards larger scales, the convective forcing on those scales also increases, because the corresponding eddies generate buoyancy anomalies by advecting the mean entropy field. The picture of a persistent small scale forcing is therefore overly simplistic in our convection driven model. } until the beta term finally becomes important. This happens when 
\begin{equation}
Ro \approx l_s H_\rho ^{-1} \Leftrightarrow \frac{l_s}{v_s} \approx \frac{k_s}{\beta_\rho},
\end{equation}
that is when the time scale of the turbulence becomes comparable to the Rossby wave time scale given by (\ref{dispersion}).
This order of magnitude estimate fails to take the anisotropic nature of Rossby waves, represented by the factor $\cos \alpha$, into account. If we speculate that the inverse cascade indeed ceases when the turbulence frequency becomes small enough to excite Rossby waves, we find by equating $v_\beta l_\beta^{-1} = v_\beta  k_\beta$ with (\ref{dispersion}) that the corresponding wave number $k_\beta$ characterizing the turbulence-wave crossover is defined by the identity 
\begin{equation}
k_\beta= \sqrt{\frac{\langle\beta_\rho\rangle}{v_\beta}\left|\cos \alpha\right|},
\label{cross-over}
\end{equation}
where $v_\beta$ denotes the amplitude of the mode $k_\beta$ in the velocity spectrum.
The wave numbers $k_\beta$ trace out a dumbbell shape in wave number space as shown in  figure \ref{dumbbell_theo}. Kinetic energy is expected to be transported from high wave numbers towards this dumbbell by the inverse cascade, which finally comes to a hold (or is at least slowed down) when $k$ reaches $k_\beta$. The minimum length scale for which Rossby waves become important is given by 
\begin{equation}
l_R = \sqrt{\frac{v_\beta}{ \langle\beta_\rho\rangle}},
\end{equation}
which, in reference to classical beta plane studies \citep{Rhines1975}, will be called the {\em compressional Rhines scale} in this paper. 

\subsection{Jet scaling}
Numerous studies of classical beta plane turbulence have revealed that the Rhines length indeed represents a good estimate for the typical jet width \citep[see e.g.][]{Vallis2006}. Therefore, we expect the same to be true for the compressional beta effect considered here. It follows that the theoretical number of jets $n_{jets}^{theo}$ should approximately follow a scaling law of the form  
\begin{equation}
n_{jets}^{theo}=C\frac{d}{l_R}=C~d\sqrt{\frac{\left \langle \beta_\rho \right \rangle }{v_\beta}} =C~d\sqrt{\frac{2\Omega \left \langle H_\rho^{-1}\right \rangle }{v_\beta}},
\label{njets}
\end{equation}
where again $d$ denotes the layer height, \alt{and}$C$ is an $O(1)$ constant \neu{and (\ref{beta_vol_avrg}) has been used to relate $\left \langle \beta_\rho \right \rangle$ to $\left \langle H_\rho^{-1}\right \rangle$}. For future reference, we express this scaling law in non-dimensional form using the scales introduced in section \ref{sec:non_dimensionalization}, 
\begin{equation}
n_{jets}^{theo}=C\sqrt{\frac{<\tilde{\beta}_\rho>}{\tilde{v_\beta}}}=C\sqrt{\frac{Pr ~ \ln \chi}{Ek ~ \tilde{v_\beta}}},
\label{njets_nondim}
\end{equation}
where \neu{the identity $\langle H_\rho^{-1} \rangle = \ln \chi / d$ is applied to express the mean inverse density scale height by the density contrast}\alt{(\ref{beta_vol_avrg}) has been used to relate $\left \langle \beta_\rho \right \rangle$ to the density contrast $\chi$}. We remind the reader that the tilde signifies non-dimensional variables in the scaling described in  section \ref{sec:non_dimensionalization}. 

\subsection{Jet formation from the potential vorticity perspective}
\label{pv_theo}
\stephan{
While we so far focussed on the compressional \neu{Rhines picture}\alt{beta mechanism} from the spectral point of view, an alternative perspective to jet formation is stirring of potential vorticity (PV), a concept dating back to the early work of \citet{Ertel1942} and \citet{Green1970}. Following \citet{Glatzmaier1981} and \citet{Glatzmaier2009} one can define the PV of a compressible fluid parcel -- appropriate for the system discussed in this paper -- as the total vorticity divided by its density, $(2\Omega + \zeta) / \rho_0$. Neglecting frictional and buoyancy effects in equation (\ref{vorticity}), this quantity can be shown to be conserved by moving fluid particles (see the appendix of \citet{Glatzmaier2009} for details),
 \begin{equation}
\frac{D}{Dt}\frac{2\Omega + \zeta}{\rho_0}=0,
\label{pv_conservation}
\end{equation}
with $D/Dt$ denoting the material derivative.  Fluid parcels moving vertically in a density-stratified layer thus have to adapt their vorticity in order to conserve potential vorticity, with sinking (rising) particles gaining \neu{positive (negative)}\alt{prograde (retrograde)} vorticity.
\\
The background density increase with depth imposes a smooth background potential density gradient onto our system. Convective turbulence, by (\ref{pv_conservation}), is expected to cause efficient PV mixing. In order to generate jets like those described in section \ref{observation}, this mixing has to be spatially non-homogenous, such that locally converging or diverging PV eddy fluxes can result.  In analogy to studies of classical beta-plane turbulence, PV staircases, i.e. stacks of well-mixed layers separated by sharp PV interfaces, may be expected (see e.g. the review by \citet{Dritschel2008}). Such staircases, by inversion, dictate jet-like mean flows.  In order to maintain the jets, the turbulent PV mixing has to be organized by the staircase in such a way as to reinforce it. As pointed out by \citet{Dritschel2008}, there is thus a certain analogy to the Philips effect that can cause staircase formation in stirred, stably stratified fluids \citep{Phillips1972,Ruddick1989,Balmforth1998}.
}

\subsection{Validity of the assumptions and error estimates}
\label{errors}
If we suppose that the jet formation dynamics involves only spatial scales below the compressional Rhines length $l_R$, then the discussion presented so far relies on the two key assumptions
\begin{eqnarray}
 \frac{\left| \beta_\rho(z) - \langle\beta_\rho\rangle \right|}{\langle\beta_\rho\rangle} &\ll& 1 \label{eq:condition2_beta_rho}, \\
\text{and} \qquad  l_R H_\rho^{-1}(z)   &\ll&1 \label{eq:condition1_beta_rho},
\end{eqnarray}
namely that $\beta_\rho$ is nearly constant with depth and that the compressional Rhines length is much smaller than the density scale height. This was argued to be true for the sample simulation 
presented in section \ref{observation} and allowed us to demonstrate the analogy to classical beta effect models on a very basic level. 

The question however remains how well the criteria (\ref{eq:condition2_beta_rho}) and (\ref{eq:condition1_beta_rho}) need to be fulfilled for the predicted effects to occur. In the next section, we present results from a suite of numerical simulations covering a wide range of control parameters. 
In order to a-posteriori assess the validity of the above assumptions in a given simulation, it is handy to express them
in terms of the \neu{global} non-dimensional control parameters describing the system, and in terms of the number of jets $n_{\text{jets}}$ observed, \neu{rather than checking the validity of (\ref{eq:condition2_beta_rho}) and (\ref{eq:condition1_beta_rho}) at all depths $z$}.

\neu{To estimate the validity of (\ref{eq:condition2_beta_rho}) and (\ref{eq:condition1_beta_rho}), $\beta_\rho(z)$ and $\langle\beta_\rho\rangle$ can be expressed in terms of $H_\rho^{-1}(z)$ according to (\ref{beta_rho}) and (\ref{beta_vol_avrg}).} From equation (\ref{back_dens}), we find that the dimensionless inverse density scale height as a function of $z$ is given by 
\begin{equation}
\tilde{H}_\rho^{-1}(z)=\frac{n\left(1-\chi^{-1/n}\right)}{\left(1 -
\left(1-\chi^{-1/n}\right) z \right)},
\label{density_scale_height_in_z}
\end{equation}
where $n$ is the polytropic index. \neu{A domain average results in}
\begin{equation}
\neu{\langle\tilde{H}_\rho^{-1}\rangle=\ln\chi,}
\label{density_scale_height_mean_nondimensional}
\end{equation}
which can also be seen directly from the definitions of $H_\rho$ and $\chi$.
\neu{In order to globally estimate the left hand side}\alt{To quantify the validity} of (\ref{eq:condition2_beta_rho}), we compute the standard deviation $\tilde{\sigma}_{\beta_\rho}$\neu{ of $\tilde{\beta}_\rho(z)$} and the\neu{ corresponding} maximum deviation $\tilde{\Delta}_{\beta_\rho,max}= \max\limits_{0 \; \le\;  z \; \le \; d} \; \; \left| \; \tilde{\beta}_\rho(z) -
\langle \tilde{\beta}_\rho \rangle \; \right|$, 
resulting in 
\begin{eqnarray}
\frac{\tilde{\sigma}_{\beta_\rho}}{\langle \tilde{\beta_\rho} \rangle} &=     & \frac{\sqrt{\frac{n^2
\left(1-\chi^{-1/n}\right)^2}{\chi^{-1/n}} - \ln^2\chi}}{\ln\chi}  \label{error_constant_H_volume}
\\
\text{and} \qquad \frac{\tilde{\Delta}_{\beta_\rho,max}}{\langle \tilde{\beta_\rho} \rangle} &= & \frac{n\left(\chi^{1/n}-1\right)
- \ln\chi}{\ln\chi}. \label{error_constant_H_max}
\end{eqnarray}
For the average and maximum error made in assuming (\ref{eq:condition1_beta_rho}), it follows
\begin{eqnarray}
\tilde{l_R} \left \langle \tilde{H}_\rho^{-1}(z) \right\rangle &\approx&  \frac{1}{n_\text{jets}} {\ln\chi} \label{error_estimate_cond_1}
\\
\text{and} \qquad \tilde{l_R} \max\limits_{z \in [0,d]} \tilde{H}_{\rho}^{-1}(z) &\approx& \frac{1}{n_\text{jets}} {n \left(\chi^{1/n}-1\right)},
\label{error_estimate_cond_2}
\end{eqnarray}
where we have used the assumption \neu{$\tilde{l}_R \approx 1 / n_\text{jets}$}\alt{$l_R \approx d / n_\text{jets}$}. 
\neu{The global quantities (\ref{error_constant_H_volume}) and (\ref{error_constant_H_max}) are domain averaged and maximum values of the left hand side of equation (\ref{eq:condition2_beta_rho}), respectively. $\beta_\rho(z)$ is almost constant with depth as long as they are much smaller than one. Moreover, the right hand sides of  (\ref{error_estimate_cond_1}) and (\ref{error_estimate_cond_2}) 
need to be much smaller than one to ensure that a typical jet scale is much smaller than the density scale height. Note that applying the global maximum values for testing the two key assumptions can ensure their validity at all depths and is thus more restrictive than using the global average values.}
The estimates (\ref{error_constant_H_volume}-\ref{error_estimate_cond_2}) will be used in section \ref{sec:moderate_to_large_density_contrasts} as indicators for the degree to which (\ref{eq:condition2_beta_rho}) and (\ref{eq:condition1_beta_rho}) hold in the simulations.

\section{\jan{Numerical simulations}}
\label{results}
In this section, results from a suite of numerical simulations are presented with the goal to test the theoretical ideas presented in section  \ref{theory}. The dimensionless form of equations (\ref{continuity} - \ref{heat}) is solved numerically for the case of a two-atomic ideal gas. The same background state and boundary conditions as described in section \ref{method} are employed. A purely conductive, stationary state is used as the initial condition. 

\subsection{Choice of parameters}
The scaling prediction (\ref{njets_nondim}) is central for the layout of our systematic parameter study. 
It suggests that extreme parameter values need to be considered in order to study the multiple jet regime. If we fix the Prandtl number to an order one value, the number of jets depends on $Ek$, $\ln \chi$ and on $v_\beta$. The latter can only be controlled in an indirect fashion by adjusting $Ra$. Since the density contrast only enters logarithmically, the Ekman number and the Rayleigh number are expected to be most influential. Our simulations reveal that $Ek \le O(10^{-7})$ is necessary in order to observe only a handful of jets. To achieve jet counts allowing to check the scaling (\ref{njets_nondim}) quantitatively, we therefore explore the Ekman number range $2.5\cdot 10^{-8} \geq Ek \geq 8\cdot 10^{-11}$. The Rayleigh number is then varied in the range $3\cdot 10^9 \leq Ra \leq 4\cdot 10^{13}$, which allows us to study a considerable range of turbulence levels. The Prandtl number, which reflects the turbulent diffusivities of entropy and momentum, is set to one. As our main goal here is to demonstrate the efficiency of the compressional beta effect, in most simulations we focus on a moderate density contrast similar to the one found in the Earth's core, $\chi=1.2$. This has the advantage that the scale of the flow structures (convective plumes, jets, etc.) are characterized by similar spatial scales at all depth, which greatly eases the interpretation of the results. 
However, since giant planets are characterized by much larger density contrasts, a smaller number of cases with $\chi=5$ and $\chi=150$ is also included in our study. A full list of the simulations carried out is given in table \ref{all_simulations}. Resolutions range from $1152^2$ to $6144 \times 7680$ grid points depending on the spatial solution structure. The numerical time step is automatically adjusted in order to fulfill the Levi-Courant stability criterion, resulting in a total number of time steps ranging from 2 to 10 million for a single simulation. 

\begin{table}
\begin{small}
\begin{tabular}{|l | r r r r |}
\multicolumn{5}{l}{$\chi=1.2$} \\
\hline
$Ek$ & $Ra$ & & & \\
\hline
$2.5\cdot 10^{-8}$  & $3\cdot 10^{9~}$, & $3.3\cdot 10^{9~}$, & $3.6\cdot 10^{9~}$, & $4\cdot 10^{9~}$, \\
                    & $5\cdot 10^{9~}$, & $6\cdot 10^{9~}$~    &                  &               \\
\hline
$8\cdot 10^{-9}$    & $2\cdot 10^{10}$, & $2.1\cdot 10^{10}$, & $2.2\cdot 10^{10}$, & $2.4\cdot 10^{10}$, \\
                    & $3\cdot 10^{10}$, & $4\cdot 10^{10}$~    &                     &                     \\
\hline
$2.5\cdot 10^{-9}$  & $9\cdot 10^{10}$, & $1\cdot 10^{11}$, & $1.1\cdot 10^{11}$, & $1.2\cdot 10^{11}$, \\
                    & $1.3\cdot 10^{11}$, & $1.4\cdot 10^{11}$, & $1.6\cdot 10^{11}$, & $3.2\cdot 10^{11}$, \\
                    & $6.4\cdot 10^{11}$, & $1.28\cdot 10^{12}$~ &                    &                    \\
\hline
$8\cdot 10^{-10}$   & $5.2\cdot 10^{11}$, & $5.3\cdot 10^{11}$, & $5.5\cdot 10^{11}$, & $6\cdot 10^{11}$, \\
                    & $7\cdot 10^{11}$, & $8\cdot 10^{11}$, & $9\cdot 10^{11}$, & $1\cdot 10^{12}$, \\
                    & $1.2\cdot 10^{12}$, & $1.6\cdot 10^{12}$, & $2.4\cdot 10^{12}$, & $4.8\cdot 10^{12}$, \\
                    & $9.6\cdot 10^{12}$~  &                    &                    &                    \\
\hline
$2.5\cdot 10^{-10}$ & $2.7\cdot 10^{12}$, & $3\cdot 10^{12}$, & $3.1\cdot 10^{12}$, & $3.2\cdot 10^{12}$, \\
                    & $3.3\cdot 10^{12}$, & $3.6\cdot 10^{12}$, & $4\cdot 10^{12}$, & $4.4\cdot 10^{12}$, \\
                    & $4.8\cdot 10^{12}$, & $1\cdot 10^{13}$,  & $2\cdot 10^{13}$, & $4\cdot 10^{13}$~  \\
\hline
$8\cdot 10^{-11}$   & $1.6\cdot 10^{13}$, & $1.7\cdot 10^{13}$, & $1.8\cdot 10^{13}$, & $2\cdot 10^{13}$, \\
                    & $2.2\cdot 10^{13}$, & $2.5\cdot 10^{13}$, & $3\cdot 10^{13}$~  &                    \\
\hline
\multicolumn{5}{l}{$\chi=5$} \\
\hline
$Ek$ & $Ra$ & & & \\
\hline
$5\cdot 10^{-8}$    & $2.75\cdot 10^{10}$, & $3\cdot 10^{10}$, & $4\cdot 10^{10}$, & $6\cdot 10^{10}$, \\
                    & $8\cdot 10^{10}$~     &                   &                   &                  \\
\hline
$5\cdot 10^{-9}$    & $8\cdot 10^{11}$, & $9\cdot 10^{11}$, & $1\cdot 10^{12}$, & $1.2\cdot 10^{12}$~ \\
\hline
\multicolumn{5}{l}{$\chi=150$} \\
\hline
$Ek$ & $Ra$ & & & \\
\hline
$1\cdot 10^{-8}$    & $2.5\cdot 10^{13}$ & & & \\
\hline
\end{tabular}
\end{small}
\caption{List of parameters for the numerical simulations presented in this paper. For all simulations, $Pr=1$ and $n=2.5$.}
\label{all_simulations}
\end{table}

\subsection{Typical time series}
\begin{figure}
\includegraphics[width=\linewidth]{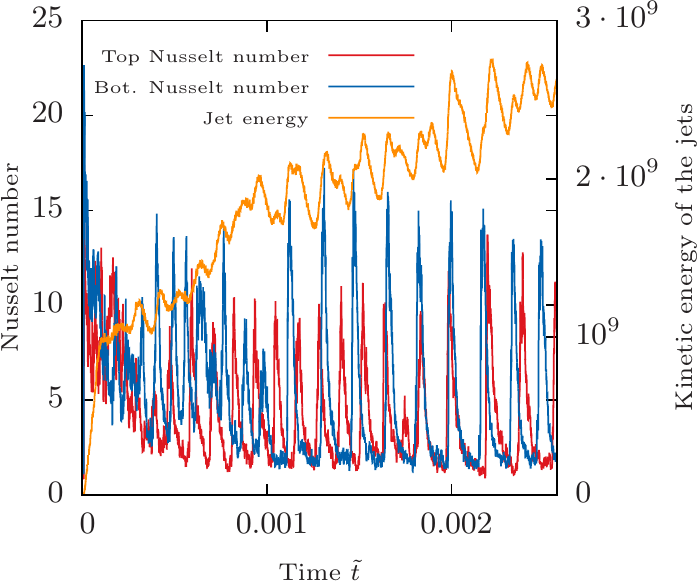}
\caption{The time series of the heat flux at the top and bottom boundaries measured in terms of Nusselt numbers is plotted for a typical simulation run. An approximate equilibrium of the fluxes, intermitted by convective bursts, is reached at about one third of the total time displayed. No quantitative analysis is performed prior to this time. The kinetic energy of the jets keeps on increasing over the whole simulation time. The parameters corresponding to the plot are \neu{$Pr=1$,} $Ek = 2.5\cdot 10^{-10}$, $Ra = 3\cdot 10^{12}$ and $\chi=1.2$. (Color version available in the online version of Icarus.)}
\label{nusselt_meanenergy}
\end{figure}

Figure \ref{nusselt_meanenergy} displays a time series from a typical simulation. Shown is the heat flux through both boundaries and the kinetic energy of the mean flow. The heat flux is given here in terms of the Nusselt number, which measures the heat flux through the respective boundary in units of the purely conductive heat flux that would be present in a non-convecting system at rest. After an initial transient, both heat fluxes fluctuate about a well defined mean value. Superimposed are longer period oscillations, which reflect the relaxation oscillations already described in section \ref{observation}. The convective bursts cause a very rapid heat flux increase, followed by a rapid decay when the shear disrupts convection. The strong variations in heat flux are accompanied by smaller variations in the mean flow energy, with both quantities exhibiting the expected phase relation.

The jet energy grows slowly on average over the entire integration time without reaching a statistically stationary equilibrium. Due to computational limitations, no attempts were made to continue the simulation over a time span sufficiently long to determine the ultimate fate of the system. The increase of the jet energy over time is not unexpected from the theory. In the absence of a \neu{scale-independent} extra drag \alt{at small wave numbers}\neu{(sometimes called bottom friction)}, the Rhines length is expected to be weakly time dependent \neu{(e.g. \citet{Danilov2004a})}, because the inverse cascade continuously  pumps energy into the large scale modes, leading to a slow increase of $l_R$ over time 
\citep{Manfroi1999,Huang2001}. However, this process tends to be very inefficient \citep{Vallis1993}, resulting in very long timescales. 

To extract quantitative averages from the simulations, the initial transient is removed by requiring that the top and bottom heat fluxes are in approximate equilibrium\footnote{As for the large density contrast $\chi=150$ the fluid flow close to the top evolves much quicker as compared to the bottom, quantitative averages had to be taken before reaching heat flux equilibrium in this particular case.} and that the kinetic energy of the mean flow exceeds that of the fluctuations. In order to represent the slow time dependence, every time series is split into a few distinct time intervals at least covering one bursting period, with averages being taken over each time interval separately. When plotting the results, values from all intervals are displayed in order to give a visual idea of the associated scatter. Note that typically, only very few merger events occur during the simulation, so that in most diagnostics, no systematic evolution is observed in subsequent time averages.

\neu{}

\subsection{Anisotropy of the kinetic energy spectrum}
\begin{figure}
\includegraphics[width=\linewidth]{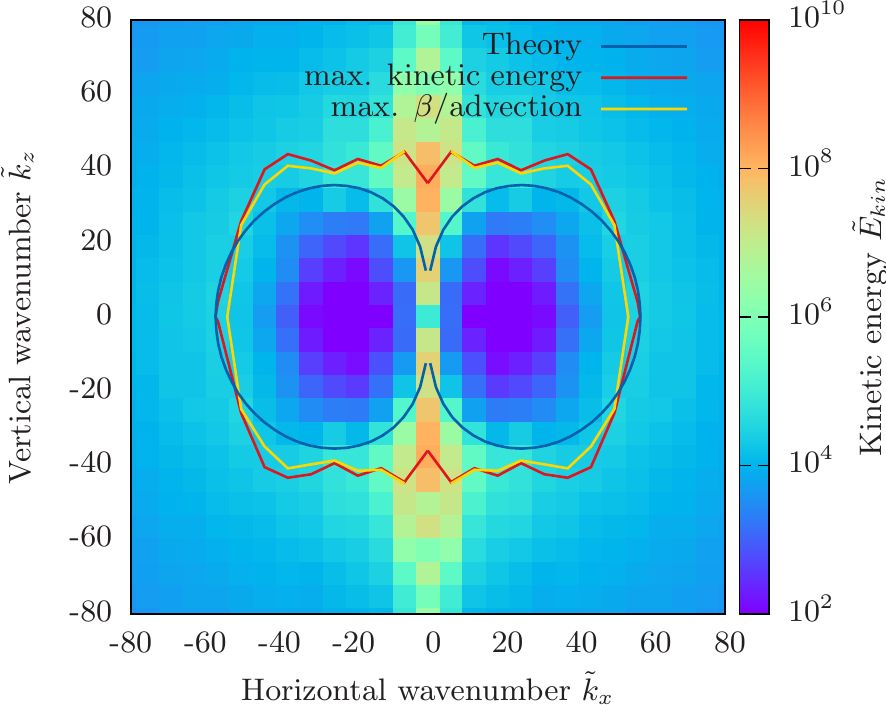}
\caption{The kinetic energy spectrum for a typical simulation run is characterized by a dumbbell shaped region that virtually holds no kinetic energy. The edge of the dumbbell  coincides with the theoretical location of the Rossby-wave-turbulence crossover plotted in blue. Furthermore the maximum of the kinetic energy (red) coincides with the maximum of the ratio of the beta- and advection terms (yellow) that are expected to control the crossover. The parameters corresponding to the energy spectrum shown are \neu{$Pr=1$,} $Ek = 2.5\cdot 10^{-10}$, $Ra = 3\cdot 10^{12}$ and $\chi=1.2$. (Color version available in the online version of Icarus.)}
\label{dumbbell}
\end{figure}

Figure \ref{dumbbell} shows the kinetic energy spectrum after the jets have formed in a typical simulation. \neu{The spectrum has been obtained by applying a raised cosine window function $0.5\{1+\cos[\pi(2\tilde{z}-1)]\}$ (e.g. \citet{Harris1978}) on the velocity field in physical space, which smoothly damps it to zero at the top and bottom boundaries.}\alt{The spectrum has been obtained by using a raised cosine window function (e.g. Harris (1978)) in the $z$-direction.} Only a small fraction of the resolved wavenumber space is shown, which eases the visual inspection of the region of interest. 

As predicted by theory the spectrum is characterized by a dumbbell shaped region holding very little kinetic energy. The kinetic energy peaks just outside this region, with a particularly significant contribution from the $\tilde{k}_x=0$ modes representing the jets. The dumbbell shape given by equation  (\ref{cross-over}), which marks the theoretical turbulence-wave crossover, is sketched in figure \ref{dumbbell} with a blue line. Note that since equation (\ref{cross-over}) is derived from simple scaling arguments, the correct prefactor determining the extent of the dumbbell is not predicted by the theory. The theoretical curve has therefore been scaled with the dominant x-wavenumber on the $\tilde{k}_z=0$ line. The shape of the theoretical curve fits the small kinetic energy region amazingly well.  

\begin{figure}
\includegraphics[width=\linewidth]{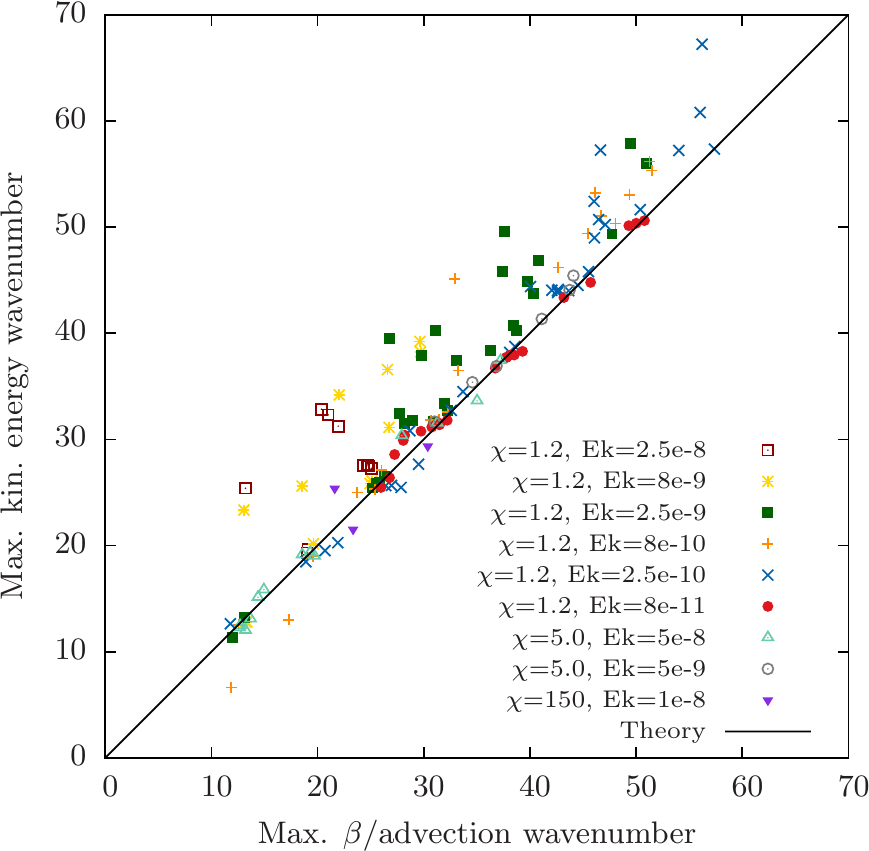}
\caption{The maximum of kinetic energy and the maximum of the ratio of beta- and advection terms on the $\tilde{k}_z=0$ line in the spectrum nearly coincide for all simulations carried out. (Color version available in the online version of Icarus.)}
\label{scaling_aposteriori}
\end{figure}
The theory largely rests on the assumption that the inverse cascade ceases when the compressional beta term in the vorticity equation reaches the amplitude of the non-linear advection term. To visualize the wave numbers relevant for this process, the yellow curve in figure \ref{dumbbell} marks the locations in wavenumber space where the spectral power of the \neu{beta}\alt{Coriolis} term divided by the spectral power of the advection term attains a maximum. Interestingly, this line nearly coincides with the location of the kinetic energy maxima. To show this, an additional red line connecting these energy maxima for each x-wavenumber is shown in figure  \ref{dumbbell}. Both curves nearly lie on top of each other. We stress that this result holds for all simulations carried out, including those assuming a large density contrast. To confirm this, \alt{we focus on the $\tilde{k}_z=0$ modes in figure \ref{scaling_aposteriori}. The}\neu{in figure \ref{scaling_aposteriori} the} x-wavenumber for which the kinetic energy attains a maximum \neu{on the $\tilde{k}_{x}$-axis} is plotted against the x-wavenumber for which the \neu{beta}\alt{Coriolis} term is strongest relative to the advection term.  Supporting our claim, all data points fall close to the diagonal\alt{ in figure \ref{scaling_aposteriori}}. 
\\
\subsection{Jet scaling}
\begin{figure}
\includegraphics[width=\linewidth]{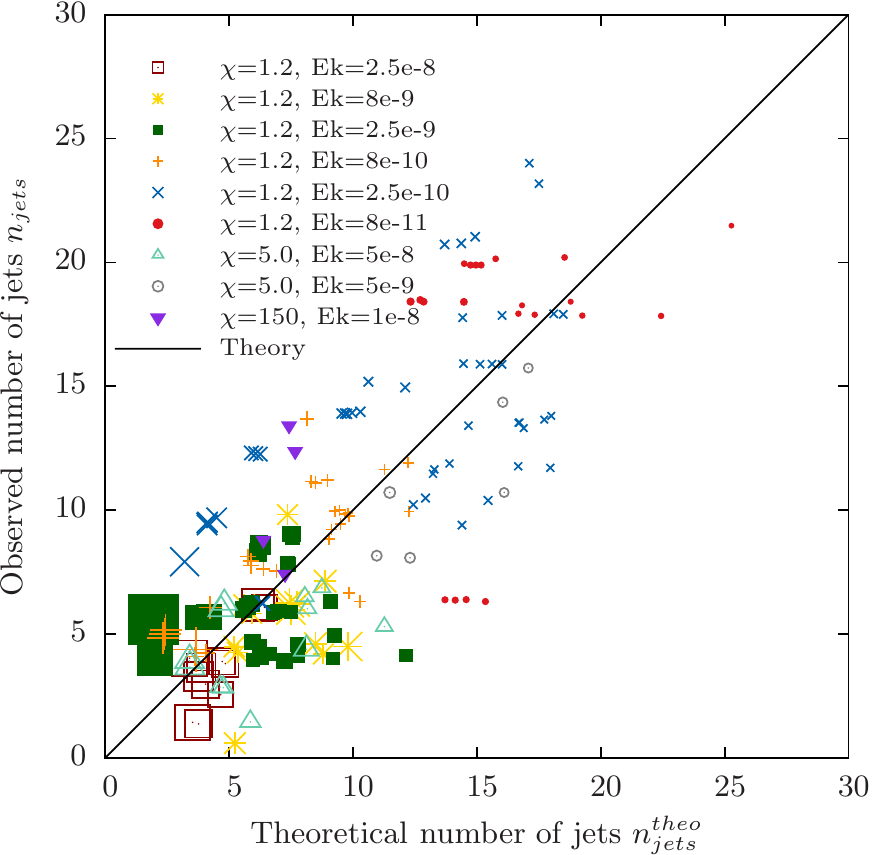}
\caption{The observed number of jets fits the theoretical predictions from the scaling law derived from the compressional Rhines scale. \neu{The symbol size is proportional to the relative importance of the advection term in comparison to the beta term, i.e. to the classical convective Rossby number $Ro_c=Ek\sqrt{Ra/Pr}$ divided by $\ln\chi$. As $Pr=1$ for all simulations, this reveals the general trend that higher $Ra$ (i.e. higher velocities) lead to fewer jets when keeping $Ek$ and $\chi$ constant. Note that for some simulations with the extreme parameters values of $Ek=2.5\cdot 10^{-10}$ and $Ek=8\cdot 10^{-11}$ few jet merging events are covered as a consequence of a relatively short time integration. As the jet velocity weakly increases over time, this leads to a constant $n_{jets}$ while $n_{jets}^{theo}$ slightly increases.} (Color version available in the online version of Icarus.)}
\label{scaling_apriori2}
\end{figure}
Of particular interest in the kinetic energy spectrum is the maximum on the $\tilde{k}_x=0$ line, because it determines the typical jet width and thus the total number of jets filling the domain. Theoretically, the number of jets is given by equation  (\ref{njets_nondim}), which relates it to the non-dimensional $\beta$-parameter $<\tilde{\beta}_\rho>= (Pr / Ek) ~ \ln\chi$ and to a typical dimensionless velocity $\tilde{v}_{\beta}$ on the jet scale, which we determine here from the kinetic energy of the peak mode in the spectrum. \neu{As in our simulations the $\tilde{k}_x=0$ modes hold most kinetic energy, in the following $\tilde{v}_{\beta}$ is taken to be the typical jet velocity $\tilde{v}_{jet}$}\alt{As in our simulations the $\tilde{k}_x=0$ modes hold most kinetic energy, in the following $\tilde{v}_{\beta}$ is taken to be the typical jet velocity, $\tilde{v}_{jet}$, which we estimate from the kinetic energy of the strongest mode in the spectrum}.\alt{ Figure \ref{scaling_apriori2} shows the number of jets obtained from the simulations versus the theoretical prediction.} The number of jets \neu{and corresponding typical jet velocity are}\alt{was} determined by interpolating the time averaged kinetic energy spectrum around the strongest mode, with the maximum of the interpolant determining the average jet number \neu{and the jet velocity}. \neu{Figure \ref{scaling_apriori2} shows the number of jets obtained from the simulations versus the theoretical prediction relying on the jet velocities for different time intervals.} Covering several orders of magnitude in $Ek$, $Ra$ and $\chi$ the plot confirms the general validity of the scaling laws (\ref{njets}) and (\ref{njets_nondim}). The best fit for the $\chi=1.2$ cases is obtained for $C=0.165$, resulting in
\begin{equation}
n_{jets}^{theo}\neu{(t)}=0.165~d\sqrt{\frac{\langle\beta_\rho\rangle}{v_{jet}\neu{(t)}}}=0.165\sqrt{\frac{Pr ~ ln \chi}{Ek ~ \tilde{v}_{jet}\neu{(t)}}}.
\label{njets_result}
\end{equation} 
\neu{The above scaling law states that the number of jets purely depends on the system depth, the beta parameter and the (time varying) jet velocity.
}

\subsection{Potential vorticity staircases}
\label{pv_results}

\begin{figure*}
\setlength{\unitlength}{\linewidth}%
\begin{picture}(1.0,0.381)
\thicklines
\put(0.0,0.0){\includegraphics[width=0.395\linewidth]{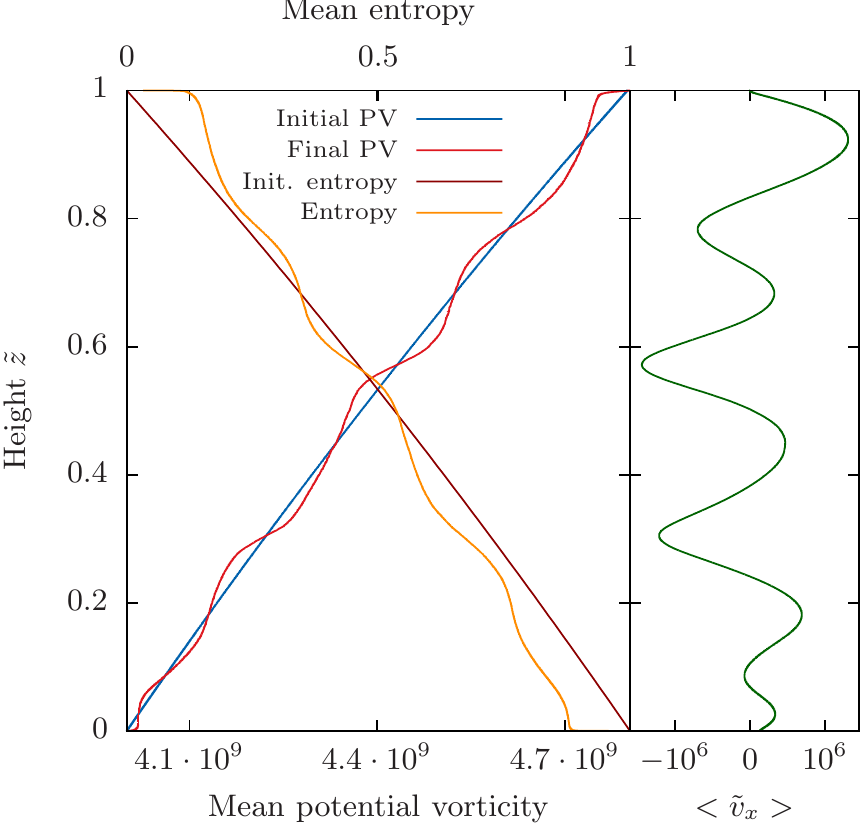}}
\put(0.4025,0.0435){\includegraphics[width=0.295\linewidth]{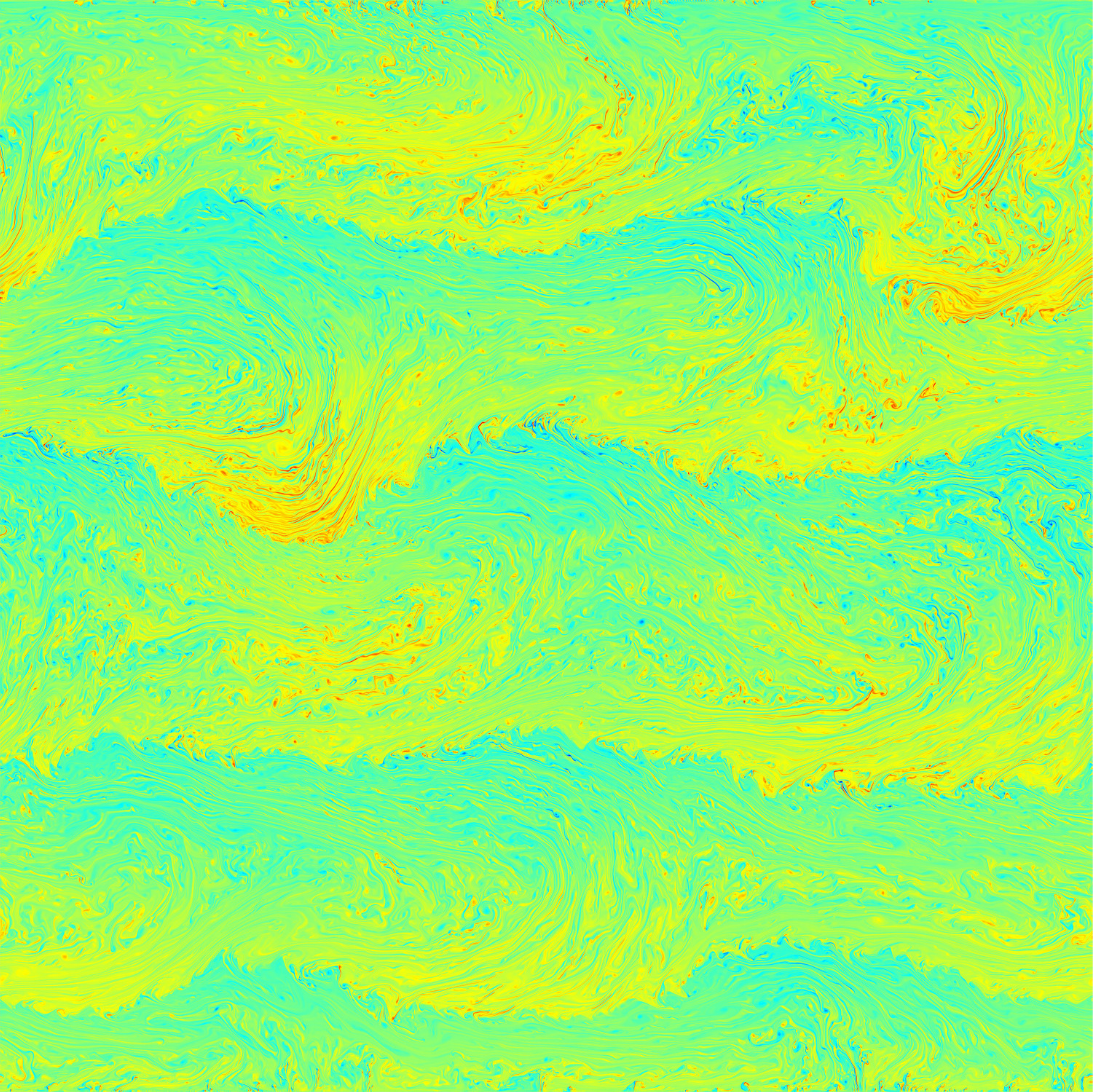}}
\put(0.705,0.0435){\includegraphics[width=0.295\linewidth]{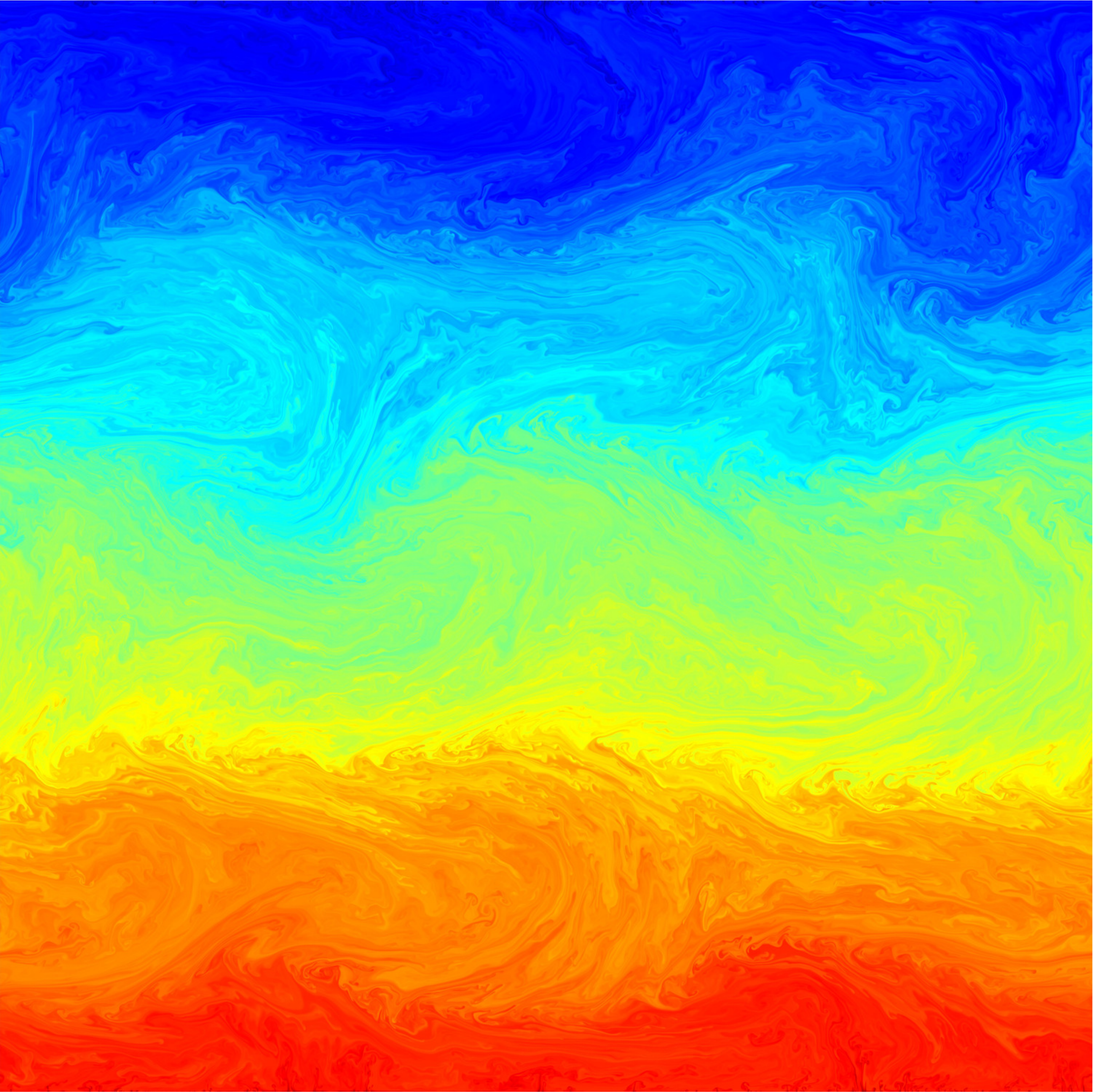}}
\put(0.4125,0.3){\fcolorbox{black}{white}{\large Vorticity $\tilde{\zeta}$}}
\put(0.715,0.3){\fcolorbox{black}{white}{\large Entropy $\tilde{s}$}}
\end{picture}
\caption{\jan{The potential vorticity (PV) staircase obtained from the simulation with the parameters \neu{$Pr=1$,} $Ek = 2.5\cdot 10^{-10}$, $Ra = 4\cdot 10^{13}$ and $\chi=1.2$ is plotted against height in red, which developed from an initial PV distribution given by the blue line. The corresponding jet velocities $<\tilde{v}_x>$ (i.e. the mean horizontal velocities) are shown in the green profile on the right. The entropy (yellow) shows the same mixing behavior as the PV \neu{(the initial/conductive entropy profile is given by the dark red line)}. Note that in contrast to the PV strong entropy gradients favor convective mixing. All profiles show temporal averages. \neu{Snapshots of the corresponding vorticity and entropy fields are shown in the middle and right panels.  In the vorticity plot, large positive values are denoted by red colors(clockwise rotation), and large negative values by blue colors (anti-clockwise rotation), with green colors depicting values close to zero. The apparent large-scale wavy structures are due to Rossby waves propagating to the right. The entropy field is shown in the rightmost panel. Entropy is directly proportional to the buoyancy of fluid parcels in the anelastic framework used for this study, with red colors denoting warm, buoyant material while blue signifies cold fluid. Eddy-transport barriers are visible in both, the vorticity- and the entropy field, located at the peak positions of the negative jet velocities.} (Color version available in the online version of Icarus.)}}
\label{pv_figure}
\end{figure*}

Figure \ref{pv_figure} illustrates that the compressional \neu{Rhines}\alt{beta} mechanism generates PV staircases  in analogy to the classical \neu{Rhines picture}\alt{beta mechanism}. For the parameters $Ek = 2.5\cdot 10^{-10}$, $Ra = 4\cdot 10^{13}$ and $\chi=1.2$, profiles of PV at the beginning and at an advanced stage of the simulation are shown. Clearly, the initial smooth PV gradient (blue), which is given by the constant planetary angular velocity $\Omega$ and the depth-dependent background density $\rho_0$, develops into a step-like PV profile (red). The corresponding mean flow is shown in green. As expected (e.g. \citet{Juckes1987} or \citet{Sommeria1989}), the sharp gradient regions  act as eddy-transport barriers and omit mixing throughout this region almost entirely. This appears to be not only due to the increased Rossby elasticity but is likely also a consequence of the strong horizontal mean flows \citep{Juckes1987}. Different from the classical \neu{Rhines picture}\alt{beta mechanism} the regions of sharp PV gradients correspond to (retrograde) westward jets, inconsistent with the classical Jovian picture associating eddy-transport barriers with prograde jets \citep{Dritschel2008}, but see \citet{Marcus2011} for a counter-example.

\stephannew{An interesting effect unique to the convection-driven case studied here is the occurrence of an unstably stratified buoyancy staircase, which acts to oppose jet formation. If viscous heating and entropy diffusion are neglected in equation (\ref{heat}), it follows that fluid particles conserve their entropy, $Ds/Dt = 0$, such that both PV and entropy must be expected to become well mixed in turbulent regions. Indeed, PV staircases are typically found to be accompanied by corresponding entropy staircases, as shown  by the orange curve in figure \ref{pv_figure}. Since buoyancy is directly proportional to entropy in the anelastic framework chosen here (see section \ref{governing_equations}), an unstably stratified buoyancy staircase results. However, in contrast to the mixing-inhibiting effect of strong interfacial PV gradients, large buoyancy gradients generally promote convective mixing across the interfaces. For the formation of zonal flows, the processes forming the PV staircase need to overcome the opposing buoyancy effects. This again illustrates that mean flow generation \neu{at lower latitudes} by a compressional \neu{Rhines mechanism}\alt{$\beta$-mechanism} \neu{-- and most likely also by a topographic Rhines mechanism --} in convective systems is far from obvious. \neu{Interestingly the opposing effects of strong PV and entropy gradients have not been discussed in previous convectively driven simulations applying a topographic beta effect \citep{Brummell1993,Jones2003,Morin2004,Heimpel2005,Morin2006,Gillet2006,Rotvig2006,Rotvig2007,Jones2009,Teed2012}.}
 }

\subsection{Moderate to large density contrasts}
\label{sec:moderate_to_large_density_contrasts}
Up to now, we have focussed on \neu{density contrasts}\alt{values of} $\chi$ only marginally exceeding one. This was motivated by the fact that the behavior then depends only  weakly on depth, which considerably simplified both the theoretical formulation and the interpretation of the numerical results.

\begin{figure}
\setlength{\unitlength}{\linewidth}
\begin{picture}(1.0,1.0)
\put(0,0){\includegraphics[width=\linewidth]{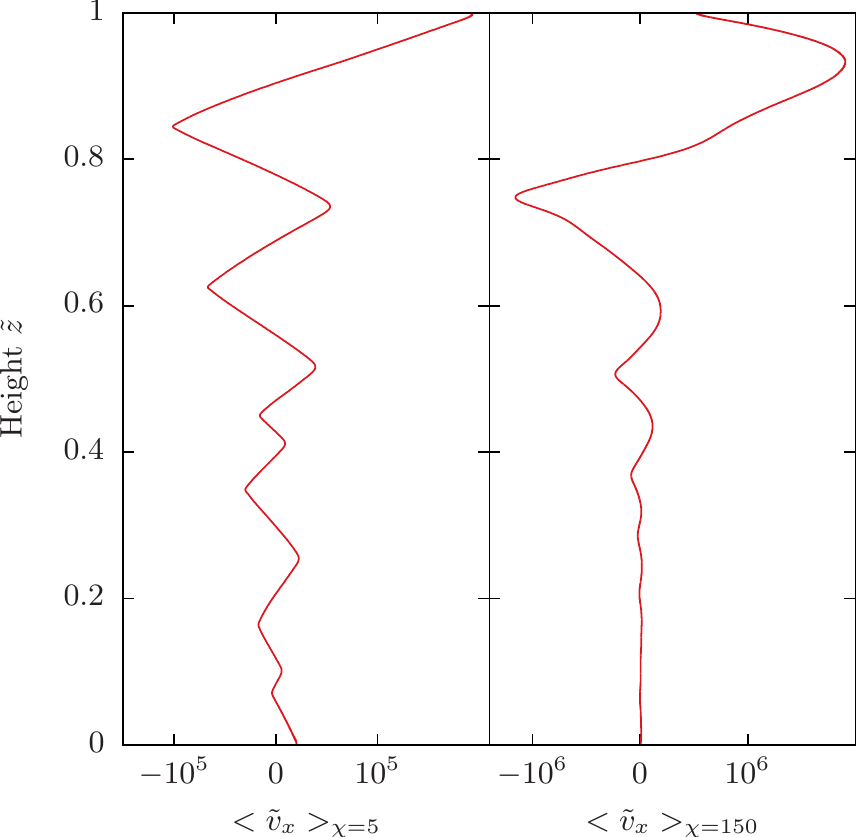}}
\put(0.16,0.905){\fcolorbox{black}{white}{$\chi=5$}}
\put(0.5885,0.905){\fcolorbox{black}{white}{$\chi=150$}}
\end{picture}
\caption{Examples of wind profiles found for large density stratifications. For all values of $\chi$ considered in this study, a multiple jet regime was found. As the density-stratification gets stronger, the vertical inhomogeneity of the system becomes increasingly noticeable, with strong jets developing at the top boundary and weak jets at the bottom. The parameters corresponding to the plots are \neu{$Pr=1$,} $Ek = 5\cdot 10^{-9}$, $Ra = 10^{12}$, $\chi=5$ and \neu{$Pr=1$,} $Ek = 10^{-8}$, $Ra = 2.5\cdot 10^{13}$, $\chi=150$, respectively.}
\label{mean_xvelos}
\end{figure}
To check how robust our findings are, we have also carried out simulations with larger density contrasts. Figure \ref{mean_xvelos} shows wind profiles obtained for $\chi=5$ and $\chi=150$. They clearly demonstrate that multiple jet patterns remain a robust feature also for strong density stratifications. Typically the jets are strongest and widest near the top of the domain, where the flow is most strongly driven, resulting in larger velocities and correspondingly in a larger {\em local} compressional Rhines scale. Note that in contrast to the jets, the convective structures are typically {\em smallest} in this region, again because convection is most strongly driven close to the top boundary. The vertical inhomogeneity tends to increase with increasing $\chi$. \neu{The wind structure shown in figure \ref{mean_xvelos} may provoke speculations that prograde flow at the surface (i.e. $\tilde{z}=1$) is preferred for pronounced density contrasts, which agrees with our experiences from other strongly density-stratified simulations. This preference may be caused by the vertically increasing Rossby-wave phase speed.}


Further support for the robustness of \neu{the compressional Rhines picture}\alt{our mechanism} comes from the fact that all simulations up to the largest density contrast $\chi=150$ follow the theoretical scaling prediction (\ref{njets_result}) for the number of generated jets.  This can be seen in figure \ref{scaling_apriori2}, which also includes the $\chi=5$ and $\chi=150$ data.  Up to about $15$ jets are observed for $\chi=5$, and all data points fall within the typical scatter exhibited by the $\chi=1.2$ cases. This clearly suggests that the theory reliably predicts the average jet scale even for large density contrasts.

\begin{figure}
\includegraphics[width=\linewidth]{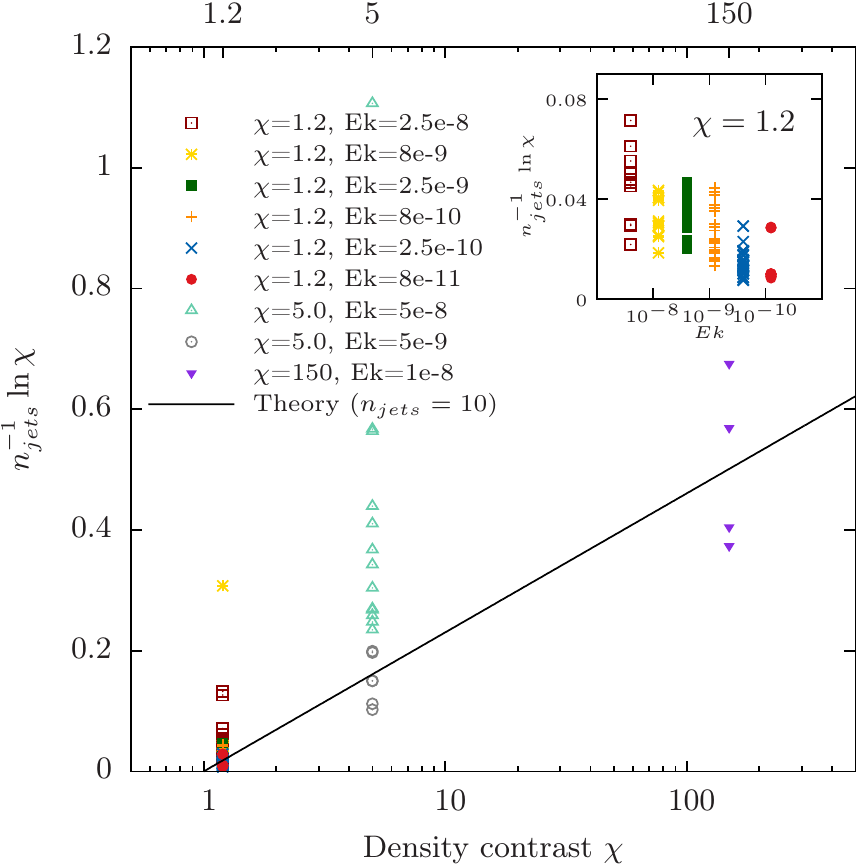}
\caption{The quantity $\tilde{l}_R \langle \tilde{H}_\rho^{-1}\rangle$, estimated as $n_{jets}^{-1} \ln\chi$ (see equation \ref{error_estimate_cond_1}), is plotted against the density contrast for all simulations carried out in this work. \neu{The black line denotes $n_{jets}^{-1} \ln\chi$ for an imaginary $n_{jets}=10$ case for comparison.} In section \ref{theory},  $\tilde{l}_R \langle \tilde{H}_\rho^{-1}\rangle$ was assumed to be much smaller than one, which seems to hold well only for small density contrasts, and only for simulations at small Ekman number, resulting in many jets. \neu{In order to clarify the Ekman number dependence in the $\chi=1.2$ case, where the points overly each other in the plot, the inset shows $n_{jets}^{-1} \ln\chi$ plotted against $Ek$ for this low density contrast. Obviously, $n_{jets}^{-1} \ln\chi$ decreases with decreasing $Ek$ as expected from the compressional Rhines scaling (\ref{njets_result}).} (Color version available in the online version of Icarus.)}
\label{figure_constraints}
\end{figure}
It is interesting in this context to reconsider the assumptions made in section \ref{theory}. In particular, in order to derive a simple theoretical scaling of the average jet thickness, we have assumed that the variations in density scale height are small across the layer, which, according to (\ref{error_constant_H_volume}) and (\ref{error_constant_H_max}), becomes questionable for large $\chi$. Furthermore, we assumed that the compressional Rhines scale is everywhere much smaller than the density scale height, which enabled us to neglect the nonlinear vorticity production and to simplify the mass conservation equation on the scales of interest.  Figure \ref{figure_constraints}, which shows an estimate of $\tilde{l}_R \langle \tilde{H}_\rho^{-1} \rangle$ based on (\ref{error_estimate_cond_1}), shows that this assumption also becomes dubious for large $\chi$, especially if we take into account that the local error may strongly exceed its \neu{domain}\alt{volume} average according to (\ref{error_estimate_cond_2}). The robustness of \stephan{the} \neu{compressional Rhines picture}\alt{mechanism} even in cases where our initial assumptions break down suggests that these should not be viewed as necessary conditions. Instead, they merely define a limit case in which the equations can be directly linked to the classical beta effect in its most basic form.

\section{Speculations about jets in planets}

\label{speculations}

A first indication for the relevance of the scaling law  (\ref{njets_result}) for planetary dynamics can be obtained by checking whether it predicts a jet scale broadly consistent with the available observations. Of course we do not expect  (\ref{njets_result}), which was derived from simple 2d numerical simulations, to accurately predict the quantitative behavior in planetary interiors. Still, it is reassuring to know that the effect under study does not lead to predictions which are off by many orders of magnitude.

We begin with the giant planets Jupiter and Saturn. In order to apply our model, we need estimates of the system depth, the average inverse density scale height and the typical jet velocities. For giant planets, the system depth is often assumed to be given by the outermost non-metallic region, where Lorentz forces can be neglected \neu{(e.g. \citet{Liu2008,Heimpel2011})}. The density profiles can then be obtained from standard planetary models.  The typical jet velocities in the equatorial plane are much harder to estimate. Motivated by the simulations of \citet{Heimpel2005,Heimpel2007} and \citet{Jones2009}, we will assume here that the Taylor Proudman theorem approximately holds on large scales, such that the jet velocities observed on the surface may provide reasonable estimates for the jet velocity in the equatorial plane. We fully acknowledge the speculative nature of this assumption. 

For Jupiter, the transition to metallic hydrogen is expected to occur at about $0.9$ planetary radii \citep{French2012}. The tangent cylinder thus   
touches Jupiter's surface at $\pm 25.8^\circ$ latitude, bounding an equatorial region characterized by jet speeds with $50 m/s \le v_{jet} \le 150m/s$, as measured on the surface by the Cassini mission \citep{Porco2003}. The remaining parameters needed are the layer depth $d=0.1r_J$, where $r_J=7\cdot10^7m$ denotes Jupiter's radius \citep{Lindal1981}, the angular velocity $\Omega=2\pi/10h$ \citep{Guillot2007} and the density contrast $\chi=5\cdot10^3$ in the top $10\%$ relative to Jupiter's $1\text{bar}$ level \citep{Nettelmann2008,French2012}.
Inserting these quantities into (\ref{njets_result}) results in $1.9\le n_{jets}^{theo}\le 3.4$ for Jupiter. This result is remarkably close to the roughly four jets observed on each hemisphere between $0^\circ$ and $\pm25.8^\circ$ latitude, \neu{bearing in mind that the scaling law was obtained from 
simple two-dimensional numerical simulations}. 

Applying the same procedure to Saturn, assuming $d=0.6r_S$ \citep{Saumon2004} with $r_S=6\cdot 10^7m$ denoting Saturn's radius \citep{Lindal1985}, $\Omega=2\pi/10h$ \citep{Guillot2007}, a density contrast of $\chi=5\cdot 10^3$ \citep{Saumon2004,Anderson2007} and choosing typical jet velocities $150 m/s \le v_{jet} \le 400 m/s$ in the region outside the tangent cylinder  \citep{Sanchez2000,Garcia2011}, we obtain $2.2\le n_{jets}^{theo} \le 3.6$. Again, this seems to be in accord with the roughy five jets observed in the regions between $0^\circ$ and $\pm 53.1^\circ$ latitude \citep{Sanchez2000,Garcia2011} bounded by the tangent cylinder on each hemisphere. 

The usual assumption that the radial extent of the jets is limited by the transition to metallic hydrogen is certainly plausible \neu{\citep{Liu2008}}, but jets extending \neu{into the gas giants' dynamo regions}\alt{deeper into the planet} can not be ruled out completely \neu{\citep{Glatzmaier2008}}. It is instructive to compute the number of predicted jets for the extreme case that the wind pattern pertains over the entire planetary radius.  For Jupiter, using $d=r_{J}$, $10m/s\le v \le =50 m/s$   (average jet velocities at higher latitudes according to \citet{Porco2003}), $\chi= 2\cdot 10^4$ \citep{Nettelmann2008,French2012}, we find $11.4\le n_{jets}^{theo} \le 25.7$. 
Similarly for Saturn, using $d=r_S$, $50m/s\le v_{jet} \le 100m/s$ \citep{Sanchez2000,Garcia2011} and $\chi = 2\cdot 10^4$ \citep{Saumon2004,Anderson2007},  $7.5\le n_{jets}^{theo} \le 10.6$ is predicted. While the process of estimating the\neu{ equatorial plane} jet velocities by typical higher latitude surface velocities is highly questionable, the results may nevertheless give a rough idea  about the expected number of jets\neu{ deep inside the planet,} in case their magnetic breaking is less effective than generally thought. If so, such jets would certainly be important in the context of planetary dynamo models \neu{(e.g. \citet{Guervilly2012})}. 

Finally, again neglecting the magnetic field, we calculate the number of jets predicted for the Earth's outer core. The outer core's angular velocity is $\Omega=2\pi/24h$ and typical large scale velocities can be inferred from the temporal behavior of the geomagnetic field, resulting in $v \approx 5\cdot 10^{-4} m/s$ \citep{Jones2007}. Together with the outer core's depth $d=2\cdot 10^6 m$ and the density contrast $\chi=1.2$, both inferred by seismology \citep{Dziewonski1981}, this leads to the prediction $n_{jets}^{theo}=54$. The minimum velocity needed to get the Earth's core out of the multiple jet regime would be $v =  1.4m/s$, which clearly seems orders of magnitude too large. Again, this result should not be interpreted as a hint that indeed a large number of jets may exists in the Earth core, because Lorentz forces will play a major role there. But it clearly illustrates that even for seemingly small density contrasts, the Boussinesq approximation, in which the \neu{compressional Rhines}\alt{compressibility beta} mechanism is \neu{excluded from the outset}\alt{eliminated}, might miss dynamically important effects.

\section{Conclusions}

\label{conclusions}

\stephan{As described in the introduction, Rhines-type \neu{dynamics}\alt{$\beta$-mechanisms} \alt{are} \neu{is} a recurring theme in theories of planetary jet dynamics, occurring in  latitudinal, topographic and compressional form, as 
 \alt{is their competition with} \neu{are} other \neu{mechanistic pictures}\alt{mechanisms} \citep[e.g.][]{Busse2002,Glatzmaier2009} \alt{relying on} \neu{that emphasize} spatial variations in the strength of the beta term. While the \stephannew{latitudinal} and the topographic beta effects have been studied extensively in the literature, in this paper we focussed on a \alt{compressional} \neu{compressional Rhines-type scenario}\alt{$\beta$-mechanism}, which has received little attention since it was originally proposed by \citet{Ingersoll1982}.  Especially in the context of the jet generation mechanism recently introduced and analyzed by  \citet{Evonuk2006,Evonuk2007,Evonuk2008,Glatzmaier2009} and \citet{Evonuk2012}, \neu{this} \alt{compressible $\beta$-mechanism} appears to be an interesting alternative \neu{view}. }

\stephannew{Indeed,}  \stephan{our study clearly demonstrates that rapidly rotating, turbulent convection has the ability to drive a multitude of deep jets by a\alt{ Rhines-type} compressional \neu{Rhines-type} \alt{$\beta$-}mechanism. Results from a suite of  two-dimensional numerical simulations employing the anelastic approximation reveal that --  as theoretically expected -- the typical jet width is well predicted by a compressional Rhines scale $l_R$, which depends on the planetary rotation rate, on the \neu{typical density scale height}\alt{amount of density increase with depth}, and on the \neu{jet velocity}\alt{vigor of the flow}. Kinetic energy spectra clearly show the anticipated  turbulence-wave crossover.

\neu{Furthermore,} potential vorticity staircases are found to develop, which are accompanied by corresponding, unstably stratified  staircases in the  \stephannew{potential density field. While the steplike distribution of potential vorticity appears to suppress the turbulent mixing across the interface regions, the buoyancy jumps over these interfaces act to facilitate convective transport.}
Such counteracting mechanisms are absent in homogenous beta-plane turbulence driven by a prescribed forcing \neu{(e.g. \citet{Vallis1993,Rhines1994,Vallis2006,Scott2012a,Scott2012b})}, which shows that the robustness of the compressional \neu{Rhines picture}\alt{$\beta$-mechanism} found in this paper is not \neu{obvious from the outset}\alt{a-priorly obvious}. The richness of the dynamics is further illustrated by the occurrence of relaxation oscillations, resulting from the interaction between the convection and the jet shear. \neu{Similar behavior has been reported in previous studies of convection in the presence of a topographic beta effect (extensive references are given in section \ref{observation}). 
}

\alt{In all cases considered, the compressional $\beta$-mechanism dominates over the Evonuk / Glatzmaier mechanism as soon as the system becomes deep enough to host multiple compressional Rhines lengths $l_R$. It thus naturally explains the occurrence of multiple jets, and operates even for a depth-independent density scale height, a condition under which the  Evonuk / Glatzmaier mechanism breaks down. In conclusion, we  expect that the compressional-beta mechanism dominates in systems deeper than $O(l_R)$, while the Evonuk / Glatzmaier mechanism takes over otherwise. Together, both pictures provide a coherent picture of the role of compressibility in generating zonal winds.}

\stephan{Whether\alt{ or not} the compressional \neu{beta effect}\alt{$\beta$-mechanism} can be expected to \neu{drive deep multiple jet patterns} \alt{contribute to deep jet generation } in planetary objects \alt{thus seems to depend} \neu{depends} on the magnitude of the  compressional Rhines lengths. A naive application to giant planets indicates that it should indeed be important in Jupiter and Saturn, where the scaling law found in this study even results in reasonable predictions for the observed number of surface jets. Somewhat more unexpected is the fact that in the absence of magnetic fields, a large number of jets is predicted \neu{also} for the Earth's outer core. Although such jets are expected to be damped by Lorentz forces, this result illustrates that the Boussinesq models typically used in this context might miss important physical effects.}

\stephan{The highly speculative nature of such applications to planetary bodies should be pointed out explicitly here. The simple model investigated in this paper was designed with the goal to demonstrate the efficiency of the compressional \neu{Rhines-type dynamics}\alt{beta mechanism} in a convecting system as simply and clearly as possible. Physical processes which are not essential for \neu{such a mechanism}\alt{this mechanism} have been intentionally neglected, irrespective of their possible importance in real systems, with the goal to study the compressional beta effect in isolation. }

\stephan{Perhaps the most important limitation of our model is that it is restricted to two-dimensional flows in an equatorial plane. It may be argued that nearly two-dimensional flows may indeed be expected there because rapid rotation tends to largely suppress flow variations along the rotation axis \neu{(e.g. \citet{Busse2002,Schaeffer2005,Calkins2012})}. 
However, the question to which degree processes \neu{only present in three dimensions} like vortex stretching and tilting can be safely neglected, and whether an inverse cascade will indeed occur, need further investigation. Encouraging in this context is the identification of an inverse cascade in numerical simulations of three-dimensional rapidly rotating Rayleigh-B\'enard convection using a reduced, asymptotic model \citep{Julien2012}. 

Different from the classical and the topographic beta effect, which produce vorticity for fluid volumes moving latitudinally or perpendicular to the rotation axis, the compressional beta effect needs radially moving fluid parcels to operate. The consequences for the dynamics in three-dimensional spherical shells have not been addressed in this paper and need further attention. }

\stephan{Another open issue is the competition with the other mechanisms proposed in the literature. }\stephannew{\neu{Processes}\alt{Effects} like the topographic or \neu{classical}\alt{latitudinal} \neu{beta effects}\alt{$\beta$-mechanisms} are absent from our model by construction.} 
\stephan{Simulations in three-dimensional spherical shells are desirable in this context. Our study however suggests that large rotation rates, corresponding to Ekman numbers below $ O(10^{-7})$, \stephannew{are} required to observe a compressional \neu{Rhines }\alt{$\beta$-}mechanism \jan{in low-latitude regions} for $O(1)$ Prandtl number fluids}.
 \stephan{Such Ekman numbers are currently not reached in numerical simulations of convection in rotating spherical shells. Among such models, only a few have included the effect of a density increase with depth so far \citep{Kaspi2009,Jones2009,Gastine2012}, typically using Ekman numbers between $O(10^{-4})$ and $O(10^{-6})$.} \neu{These highly resolved 3-d simulations as well as their incompressible counterparts (e.g. \citet{Heimpel2005,Heimpel2007}) consistently only find two jets on the equatorial plane.}

\stephannew{An interesting first step towards identifying a compressional \neu{Rhines }\alt{$\beta$-}mechanism in rotating spherical shells has recently been carried out in parallel to our study by \citet{Gastine2014}. Although their work is restricted to moderate Ekman numbers, the authors argue that the typical length scales for high latitude jets are better described by a compressional Rhines scale than by its topographic counterpart. Both effects are however difficult to separate, because the expected scaling difference is small in the accessible parameter range. The next generation of numerical convection models  will hopefully allow us to enter the relevant low Ekman number regime, such that the relative strength and mutual interaction of the compressional beta effect and other relevant mechanisms can be studied in detail. Until then, the study presented in this paper may serve at least as clear and unambiguous evidence for the general ability of \neu{the}\alt{a Rhines-type,} compressional \neu{beta effect}\alt{$\beta$-mechanism} to \neu{generate}\alt{drive} multiple zonal jets in convective systems.}

\jan{Is the compressional beta effect indeed a source of planetary zonal winds? Although we cannot answer this question here, the main conclusion of our study is that it should be seen as one of several possible candidates involved in driving (deep) planetary jet patterns. }

\section*{Acknowledgements}
\stephannew{We thank Thomas Gastine for discussions and for making a manuscript describing his recent work available to us.}
The computations have been carried out on the PALMA computer cluster \stephannew{at} M\"unster University and on the super computer JUQUEEN at the Forschungszentrum J\"ulich. This work was supported by the the German Science Foundation under the Priority Program 1488  (Planetary Magnetism).


\bibliographystyle{elsarticle-harv}
\bibliography{Literature}







\end{document}